\documentclass[aps,pre,twocolumn,showpacs,superscriptaddress,groupedaddress]{revtex4-1}
\usepackage{graphicx}
\usepackage{textcomp}
\usepackage{amsmath}
\usepackage{commath}
\usepackage{color}

\begin{document}
	 \title{
		Doubly-nonlinear waveguides with self-switching functionality selection capabilities}
\author{Weijian Jiao, Stefano Gonella}
\affiliation{Department of Civil, Environmental, and Geo- Engineering\\University of Minnesota, Minneapolis, MN 55455, USA\\}

\begin{abstract}
In this article, we investigate the effects of the interplay between quadratic and cubic nonlinearities on the propagation of elastic waves in periodic waveguides. Through this framework, we unveil an array of wave control strategies that are intrinsically available in the response of doubly-nonlinear systems and we infer some basic design principles for tunable elastic metamaterials. The objective is to simultaneously account for two sources of nonlinearity that are responsible for distinct and complementary phenomena and whose effects are therefore typically discussed separately in the literature. Our study explicitly targets the intertwined effects that the two types of nonlinearity exert on each other, which modify the way in which their respective signatures are observed in the dynamic response. Through two illustrative examples we show how the dispersion correction caused by cubic nonlinearity can be used as an internal switch, or mode selector, capable of tuning on/off certain high-frequency response features that are generated through quadratic mechanisms. 
To this end, a multiple scale analysis is employed to obtain a full analytical solution for the nonlinear response that includes a complete description of the dual frequency-wavenumber dispersion correction shifts induced on all the branches, and elucidates the conditions necessary for the establishment of phase matching condition.
\vspace{0.4cm}
\end{abstract}

\maketitle

\section{Introduction}
The most distinctive and fundamental property of phononic crystals (and periodic structures at large) is their inherent ability to attenuate waves whose spectral content falls within frequency intervals known as phononic bandgaps \cite{Hussein_2014}. This property stems directly from the periodicity of the material and strictly requires spatial compatibility between the wavelength of the excitation and the characteristic size of the unit cell. By incorporating internal resonators in the unit cells, bandgaps with subwavelength characteristics can be opened at low frequencies \cite{Liu_2000}, with onsets at the natural frequencies of the resonators. These gaps survive the relaxation of periodicity and can therefore be observed in spatially randomized systems \cite{Celli_2015}. A medium with locally-resonant mechanisms displays negative effective mass density \cite{Huang_2009}, which has also been exploited, in conjunction with other negative effective moduli, to achieve negative refraction \cite{Huang_2014}. 
In addition to modifying the bandgap spectrum, the introduction of an auxiliary microstructure of resonators enriches the band structure of a crystal with new wave modes characterized by deformation mechanisms and dispersive properties that are germane to the resonators motion, thus stretching its functionality landscape.


While the bulk of the literature on phononic crystals has been concerned with the linear regime, recently, nonlinearity has received considerable attention as a mean to endow materials with unprecedented tunability. The main manifestation of nonlinearity is the amplitude dependence of the dynamic response, which offers additional opportunities to tune the spectro-spatial characteristics of a periodic structure by simply controlling the amplitude of excitation. 

The weakly nonlinear regime has been studied with special focus on cubic and quadratic nonlinearity, due to their tractability and to the fact that they represent the dominant terms in many arbitrary nonlinear material behaviors. The main manifestation of cubic nonlinearity is a frequency correction of the dispersion bands, which can be predicted by perturbation methods, as reported in classical applied mathematics textbooks \cite{Holmes}. Narisetti et al. \cite{Narisetti_2010, Narisetti_2011} revisited this concept and updated the perturbation technique in the context of periodic structures to show amplitude dependence of the dispersion branches of phononic media, while experimental evidence of the frequency shifts achievable in granular chains was reported in \cite{Cabaret_2012, Bonanomi_2015}. Other interesting cubic nonlinear phenomena, not strictly limited to weak nonlinearity, include  intrinsic localized modes \cite{Sievers_1988}, chaotic bands \cite{Wen_2017}, and third harmonic generation (THG) \cite{Xin_2018}. In addition, Moore et al. \cite{Moore_2018_PRE} achieved nonreciprocity in coupled oscillators featuring strong cubic nonlinearity realized using thin acrylic strips under transverse deformation.

In the case of quadratic nonlinearity, the main nonlinear effect of interest is the second harmonic generation (SHG) \cite{Nayfeh, Zarembo_1971, Hamilton}, while the correction of the dispersion relation has been shown to be negligible \cite{Ganesh_2013}. Along these lines, perturbation analysis has been used to find the explicit solution of second harmonic terms in monoatomic granular chains, and to explore the interplay of SHG and dispersion \cite{Tournat_2013}. Recently, Mehrem et al. \cite{Mehrem_2017} have conducted a series of experiments to confirm the existence of SHG in a nonlinear chain consisting of repulsive magnets. A few recent works have addressed the problem of nonlinear wave propagation in periodic structures with multiple degrees of freedom, where the availability of complex dispersive characteristics results in the nonlinear enrichment of the default linear response and in the development of new wave manipulation capabilities. Among these effects, we recall modal mixing in granular systems \cite{Ganesh_PRL_2015, wallen2017} and augmented spatial directivity in $2$D metamaterial lattices \cite{Ganesh_2017}. Jiao and Gonella \cite{JIAO2018} incorporated locally resonant mechanisms in nonlinear waveguides to realize via SHG inter-modal energy tunneling from a fundamental flexural mode to a nonlinearly-activated axial mode with subwavelength characteristics, and subsequently confirmed this phenomenon experimentally \cite{Jiao_2018_PRA}. Another example of this quadratic nonlinear tunneling is an acoustic rectifier realized by coupling a superlattice with a nonlinear microbubble suspension, in which a rectified energy flux of acoustic waves can be nonlinearly generated \cite{Liang_2010}. Moreover, phase matching condition (PMC) can be established when the fundamental harmonic and the second harmonic have the same phase velocity, resulting in cumulative second harmonic propagation \cite{Konotop_1996, Hamilton_2003}. This unique feature significantly increases the strength of SHG, thus providing a practical way for experimental verification \cite{deng_2005, Bermes_2007} and applications of nonlinear effects \cite{Jiao_2018_PRA, Deng_2007_APL}.

While, in most relevant contexts, the effects of quadratic and cubic nonlinearity have been studied separately, a few works have dealt with these two types of nonlinearity jointly to study wave propagation in classical anharmonic lattices \cite{Huang_1993, Askar_1973, Huang_1998} and in granular systems obeying Hertz contact law \cite{Cabaret_2012, Wallen_2017_PRE}. However, little attention has been paid to the implications of the quadratic-cubic (Q-C) interplay for wave control and tunability. The scope of this article is precisely to fill this gap. To this end, we consider a medium - here a spring-mass chain on nonlinear elastic foundation - that features simultaneously both sources of nonlinearity (at different orders) embedded in the chain and in the support elastic elements. The coexistence of Q-C effects opens unprecedented opportunities for amplitude-controlled wave tuning. We are interested in how the cubic nonlinearity can be used to effectively modify the way in which the quadratic nonlinear effects manifest in the response. Specifically, we investigate whether the dispersion correction caused by cubic nonlinearity can be used as an internal switch (or mode selector) capable of switching on/off certain high-frequency features of the response that are generated by the quadratic mechanisms, or at least controlling their strength. This additional tuning capability is fully intrinsic to the system, as the selection of the functionality is solely due to the nonlinear effects, and therefore controlled by the amplitude of excitation, without resorting to any active external tuning.

In section 2, a multiple scales analysis is applied to obtain the full nonlinear response of the chain. 
Due to the Q-C interaction, SHG is observed with some unique modal characteristics, including magnified dispersion shift and tunable PMCs. In section 3, we proceed to verify our analytical model by performing a suite of numerical simulations with absorbing boundary layers. Finally, in section 4, two examples of tunable systems with self-adaptive switches are given to illustrate the wealth of wave manipulation capabilities that directly leverage the Q-C nonlinear interaction. The highlights of this investigation are summarized in the concluding section 5.

\section{Multiple scales analysis}

\subsection{Governing equations and linear dispersion relation}

\begin{figure} [!htb]
	\centering
	\includegraphics[scale=0.35]{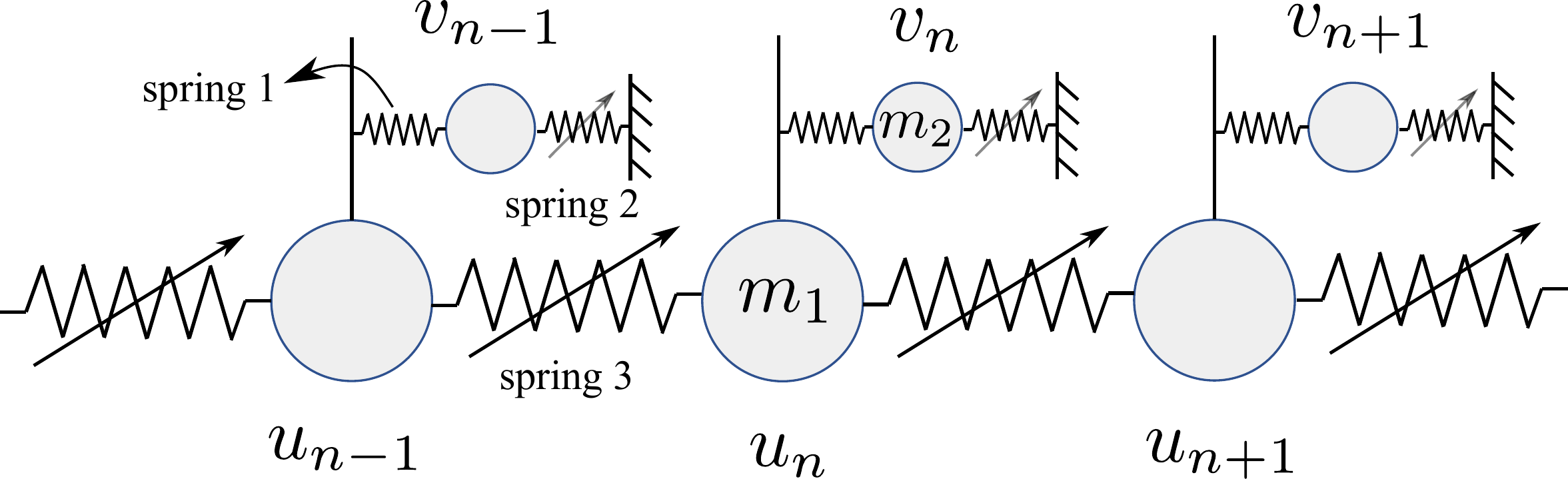}
	\caption{Nonlinear spring-mass chain with locally-resonant nonlinear elastic foundations.}
	\label{sketch}
\end{figure}

Our reference system is the spring-mass chain with a locally-resonant nonlinear foundation shown in Fig.~\ref{sketch}.
In each cell, the primary mass $m_1$ is connected to an auxiliary mass $m_2$ through linear spring $1$, and $m_2$ is connected to a fixed point via spring $2$, which is assumed to feature cubic nonlinearity, thus implementing a locally-resonant nonlinear foundation. Spring $3$, connecting $m_1$ to its neighbors along the chain, is assumed to feature quadratic nonlinearity. Under weak nonlinearity assumptions, the force-stretch relations in the three springs can be expressed as:
\begin{align}\label{Restoring force}
\begin{split}
f_1&=J_2\delta \\
f_2&=G_2\delta+\epsilon_1 G_4\delta^3 \\
f_3&=K_2\delta+\epsilon_2 K_3\delta^2
\end{split}
\end{align}
where $\epsilon_1$ and $\epsilon_2$ are small parameters for the cubic and quadratic nonlinearities, respectively. In this analysis, we assume that $\epsilon_1>\epsilon_2$ in order to establish the hierarchy between nonlinear terms necessary to achieve the desired tuning effects.

The equations of motion for the two masses ($m_1$ and $m_2$) in $n^{th}$ unit cell can be written in matrix form as:
\begin{widetext}
 \begin{equation}\label{governing_eqns}
 \begin{split}
&\begin{bmatrix} m_1 & 0 \\ 0 & m_2 \end{bmatrix} \begin{Bmatrix}\ddot{u}_n \\ \ddot{v}_n\\ \end{Bmatrix} +
\begin{bmatrix} 2K_2+J_2 & -J_2 \\ -J_2 & J_2+G_2 \end{bmatrix} \begin{Bmatrix}u_n \\ v_n\\ \end{Bmatrix} +
\begin{bmatrix} -K_2 & 0 \\ 0 & 0 \end{bmatrix} \begin{Bmatrix}u_{n-1} \\ v_{n-1}\\ \end{Bmatrix} +
\begin{bmatrix} -K_2 & 0 \\ 0 & 0 \end{bmatrix} \begin{Bmatrix}u_{n+1} \\ v_{n+1}\\ \end{Bmatrix} \\
&+ \epsilon_1\begin{Bmatrix} 0 \\ G_4 v_n^3\\ \end{Bmatrix} + \epsilon_2\begin{Bmatrix} K_3\left[ \left(u_n-u_{n-1} \right) ^2 - \left(u_{n+1}-u_n \right) ^2\right]  \\ 0\\ \end{Bmatrix}= \begin{Bmatrix}0 \\ 0\\ \end{Bmatrix}
\end{split}
\end{equation}
\end{widetext}
In the spirit of multiple scales analysis, we introduce a fast spatio-temporal variable $\theta_n=\xi n-\omega t$,  where $\xi$ and $\omega$ are the normalized wavenumber and frequency, respectively, and $n$ is an integer mass index, as well as the two slow variables $s=\epsilon_1 n$ (spatial), and $\tau=\epsilon_1 t$ (temporal). Accordingly, the cell nodal response is expressed as an expansion, up to $O(\epsilon_2)$, in successive approximations as:
\begin{align}\label{soln_assumed}
\begin{split}
\mathbf{u}_n&=\begin{Bmatrix}u_{n}(t) \\ v_{n}(t)\\ \end{Bmatrix}\\
&=\begin{Bmatrix}u^0_n(\theta_n,s,\tau) +\epsilon_1u^1_n(\theta_n,s,\tau) +\epsilon_2u^2_n(\theta_n,s,\tau) \\ v^0_n(\theta_n,s,\tau) +\epsilon_1v^1_n(\theta_n,s,\tau) +\epsilon_2v^2_n(\theta_n,s,\tau) \\ \end{Bmatrix}
\end{split}
\end{align}
Substituting this into Eq.~\ref{governing_eqns} and using the chain rule to resolve the time derivatives, one obtains the following system of cascading linear equations at each order of expansion:
\begin{align}
\begin{split}\label{order1}
O(1)&:\quad \omega^2\mathbf{M}\frac{\partial^2 \mathbf{u}^0_n}{\partial \theta_n^2}+\mathbf{K}_1\mathbf{u}_n+\mathbf{K}_2\mathbf{u}^0_{n-1}+\mathbf{K}_2\mathbf{u}^0_{n+1}=\mathbf{0}
\end{split}
\\
\begin{split}\label{order2}
O(\epsilon_1)&:\quad \omega^2\mathbf{M}\frac{\partial^2 \mathbf{u}^1_n}{\partial \theta_n^2}+\mathbf{K}_1\mathbf{u}^1_n+\mathbf{K}_2\mathbf{u}^1_{n-1}+\mathbf{K}_2\mathbf{u}^1_{n+1}=\mathbf{f}^1
\end{split}
\\
\begin{split}\label{order3}
O(\epsilon_2)&:\quad \omega^2\mathbf{M}\frac{\partial^2 \mathbf{u}^2_n}{\partial \theta_n^2}+\mathbf{K}_1\mathbf{u}^2_n+\mathbf{K}_2\mathbf{u}^2_{n-1}+\mathbf{K}_2\mathbf{u}^2_{n+1}=\mathbf{f}^2
\end{split}
\end{align}
where $\mathbf{M}=\begin{bmatrix} m_1 & 0 \\ 0 & m_2 \end{bmatrix}$; $\mathbf{K}_1=\begin{bmatrix} 2K_2+J_2 & -J_2 \\ -J_2 & J_2+G_2 \end{bmatrix} $; $\mathbf{K}_2=\begin{bmatrix} -K_2 & 0 \\ 0 & 0 \end{bmatrix} $ and the forcing terms at  $O(\epsilon_1)$ and $O(\epsilon_2)$ are:
\begin{align}
\begin{split}\label{forcef1}
\mathbf{f}^1&=2\omega \mathbf{M}\frac{\partial^2 \mathbf{u}^0_n}{\partial \theta_n \partial \tau}\\
&-\begin{Bmatrix} 0 \\ G_4 \left( v^0_n\right) ^3\\ \end{Bmatrix}+\begin{Bmatrix} K_2\left(\frac{\partial u^0_{n+1}}{\partial s}- \frac{\partial u^0_{n-1}}{\partial s} \right)\\ 0\\ \end{Bmatrix}
\end{split}\\
\begin{split}\label{forcef2}
\mathbf{f}^2&=-\begin{Bmatrix} K_3\left[ \left(u^0_n-u^0_{n-1} \right) ^2 - \left(u^0_{n+1}-u^0_n \right) ^2\right]  \\ 0\\ \end{Bmatrix}
\end{split}
\end{align}

The general solution at $O(1)$ can be expressed as 
\begin{equation} \label{soln1_O1}
\mathbf{u}_n^0=A(s,\tau) \bm{\phi} e^{i\theta_n}+A^*(s,\tau)  \boldsymbol{\phi}^* e^{-i\theta_n}
\end{equation}
where $A$ is an amplitude term that only depends on the slow scale variables $(s,\tau)$, $\boldsymbol{\phi}=\begin{Bmatrix} \phi_u \\ \phi_v\\ \end{Bmatrix}$ is a modal vector, and $(\cdot)^*$ denotes the complex conjugate of a variable. Imposing Bloch conditions on the fast scale variable $\theta_n$ between neighboring cells, we obtain
\begin{equation}\label{soln2_O1}
\mathbf{u}_{n\pm 1}=A(s,\tau) \bm{\phi} e^{i\theta_n}e^{\pm i\xi}+A^*(s,\tau) \boldsymbol{\phi}^* e^{-i\theta_n}e^{\mp i\xi}
\end{equation}
Substituting Eq.~\ref{soln1_O1} and Eq.~\ref{soln2_O1} into Eq.~\ref{order1}, the linear dispersion relation is obtained by solving the eigenvalue problem
\begin{equation}\label{eigenvaule_prob}
\left( -\omega^2 \mathbf{M}+\mathbf{K}(\xi)\right) \boldsymbol{\phi}=\mathbf{0}
\end{equation}
where $\mathbf{K(\xi)}=\begin{bmatrix} 2K_2\left( 1-\cos\xi\right) +J_2 & -J_2 \\ -J_2 & J_2+G_2 \end{bmatrix}$. Two branches, denoted as ``L" for the lower branch and ``U" for the upper branch, are obtained analytically as:
\begin{equation}\label{Linear_dispersion}
\omega_{L/U}=\sqrt{\frac{b\mp\Delta(\xi)}{2a}}
\end{equation}
where $\Delta(\xi)=\sqrt{b^2-4ac}$, with $a=m_1m_2$, $b=\left( m_1+m_2\right) J_2+2m_2K_2\left(1-\cos \xi \right)+m_1G_2$, and \\
$c=2K_2\left(J_2+G_2 \right)\left(1-\cos \xi \right)+J_2G_2.$
The amplitude ratio for each mode is given by
\begin{equation}\label{modal_vector}
\boldsymbol{\phi}_{L/U}=\begin{Bmatrix}\frac{-\omega^2_{L/H}m_2+J_2+G_2}{J_2} \\ 1\ \end{Bmatrix}
\end{equation}

\begin{figure} [!htb]
	\centering
	\includegraphics[scale=0.6]{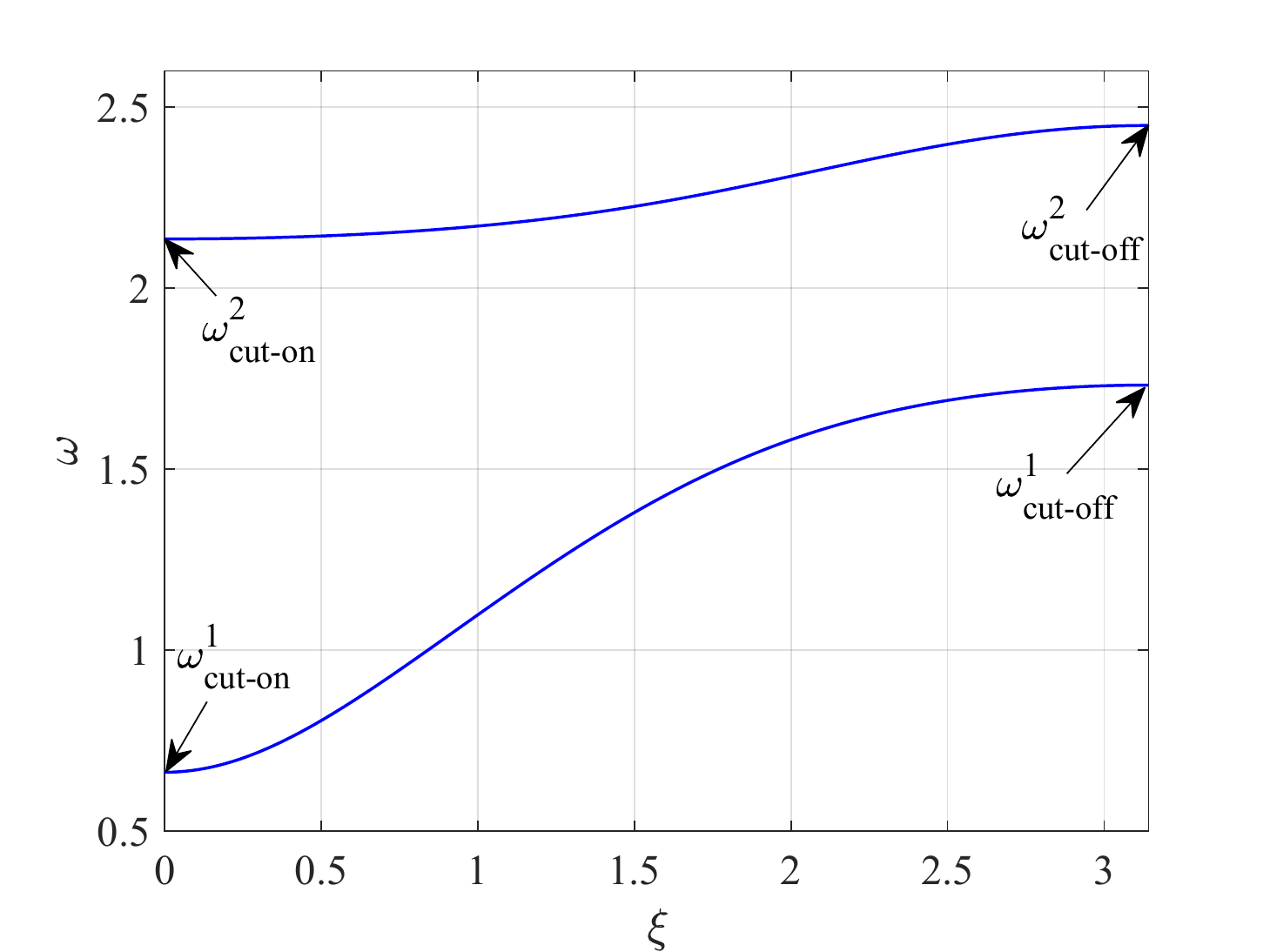}
	\caption{Linear dispersion relation for the chain with elastic foundation.}
	\label{Linear_band}
\end{figure}

Fig.~\ref{Linear_band} depicts the dispersion relation of the linearized model obtained for an arbitrary choice of parameters: $m_1=1$, $m_2=0.5$, $K_2=J_2=G_2=1$ (standard SI units are adopted throughout the paper and omitted for simplicity). Visual inspection of the plot reveals a few peculiar dispersive characteristics of the chain. (1) The system is fully gapped at low frequencies, between $0$ and the cut-on frequency for the first branch $\omega^1_\mathrm{{cut-on}}$, as expected for a chain on an elastic foundation. (2) A locally-resonant bandgap is opened in the interval between $\omega^1_\mathrm{{cut-off}}$ and $\omega^2_\mathrm{{cut-on}}$. The positive group velocity exhibited by the upper branch at $\xi=0$ is another clear signature of a locally-resonant mechanism. 

\subsection{Nonlinear correction of the dispersion relation}

We now proceed to solve the equations at $O(\epsilon_1)$. Following standard modal decomposition \cite{Meiro}, we introduce the modal coordinates $\mathbf{z}_n$, such that $\mathbf{u}^1_n=\boldsymbol{\Phi} \mathbf{z}_n$, where $\boldsymbol{\Phi}$ is the modal matrix (featuring the modal vectors $\boldsymbol{\phi}$ on the columns). Pre-multiplying Eq.~\ref{order2} by $\boldsymbol{\phi}^H$, where $(\cdot)^H$ denotes conjugate transpose, yields
\begin{equation}\label{order1_decoupled}
\omega^2 \bar{m}\frac{\partial^2 z_n}{\partial \theta_n^2}+\bar{k}z_n=\boldsymbol{\phi}^H \mathbf{f}^1
\end{equation}
where $\bar{m}=\boldsymbol{\phi}^H \mathbf{M}\boldsymbol{\phi}$ and $\bar{k}=\boldsymbol{\phi}^H \mathbf{K}\boldsymbol{\phi}$ are the modal mass and stiffness matrices. To prevent unbounded solution, the secular terms in $\boldsymbol{\phi}^H \mathbf{f}^1$ need to be eliminated \cite{MANKTELOW_2014}, which, upon some algebraic steps, leads to the required condition
\begin{equation}\label{secularity_eqn}
\frac{\partial A}{\partial \tau}+\lambda \frac{\partial A}{\partial s}+i\mu \abs{A}^2 A=0
\end{equation}
where $\lambda=\frac{K_2 \abs{\phi_u}^2 \sin \xi}{\omega \bar{m}}$ and $\mu=\frac{3G_4 \abs{\phi_v}^4}{2\omega \bar{m}}$. Note that Eq.~\ref{secularity_eqn} is a first-order PDE that controls the evolution of $A$ with respect to slow space and time variables. In other words, eliminating secularity from the solution results in a constraint on the amplitude of the fundamental harmonic. The amplitude $A$ can be written in polar form as
\begin{equation}\label{Amplitude}
A\left(s,\tau \right) = \alpha\left(s,\tau \right)e^{-i\beta\left(s,\tau \right)}
\end{equation}
Substituting it into Eq.~\ref{secularity_eqn}, yields a complex-variable algebraic equation, the soution of which requires that real and imaginary parts vanish individually, yielding the two conditions: 
\begin{align}\label{Amplitude_eqns}
\begin{split}
\frac{\partial \alpha}{\partial \tau}+\lambda \frac{\partial \alpha}{\partial s}&=0
\\
\frac{\partial \beta}{\partial \tau}+\lambda \frac{\partial \beta}{\partial s}&=\mu \alpha^2
\end{split}
\end{align}
The general solutions for $\alpha$ and $\beta$ are 
 \begin{align}\label{Amplitude_soln}
 \begin{split}
\alpha&=\alpha_0\left(s-\lambda \tau \right) \\
\beta&=\beta_0\left(s-\lambda \tau \right) + \beta^*
 \end{split}
 \end{align}
The full expression for $\beta$ contains a particular solution $\beta^*$, which can be expressed either in terms of variable $\tau$ as $\beta^*=\mu \alpha^2 \tau$, or in terms of variable $s$ as $\beta^*=\mu \alpha^2 s/\lambda$. Accordingly, in order to determine $\beta^*$, either initial conditions or boundary conditions must be supplied, based on the problem of interest. For example, by imposing the initial harmonic amplitude profile $\mathbf{u}_n=A_0 \boldsymbol{\phi}\sin \xi n$ at $t=0$, it follows that $\alpha=A_0$, $\beta_0=0$ and $\beta^*=\mu \alpha^2 \tau$. Thus, the fundamental solution at $O(1)$ is
\begin{equation}\label{modified_soln1_O1}
\mathbf{u}_n^0=A_0 \bm{\phi} e^{i\left[ \xi n-\left(\omega + \epsilon_1 \mu A^2_0 \right) t\right] }+c.c.
\end{equation}
where $c.c.$ denotes the complex conjugate of all preceding terms. Eq.~\ref{modified_soln1_O1} can be seen as a harmonic wave with wavenumber $\xi$ and frequency shifted by the correction term $\epsilon_1 \mu A^2_0$. In contrast, if the harmonic displacement time history $\mathbf{u}_n=A_0 \boldsymbol{\phi}\sin \omega t$ is imposed at one end of the chain (e.g., at $n=0$), it follows that $\alpha=A_0$, $\beta_0=0$ and $\beta^*=\mu \alpha^2 s/\lambda$. With this, the fundamental solution reads
\begin{equation}\label{modified_soln2_O1}
	\mathbf{u}_n^0=A_0 \bm{\phi} e^{i\left[ \left( \xi-\epsilon_1 \mu A^2_0/\lambda \right) n-\omega t\right] }+c.c.
\end{equation}
in which the term $\epsilon_1 \mu A^2_0/\lambda$ is a correction shift for the wavenumber.
 \begin{figure*} [!htb]
	\centering
	\includegraphics[scale=0.6]{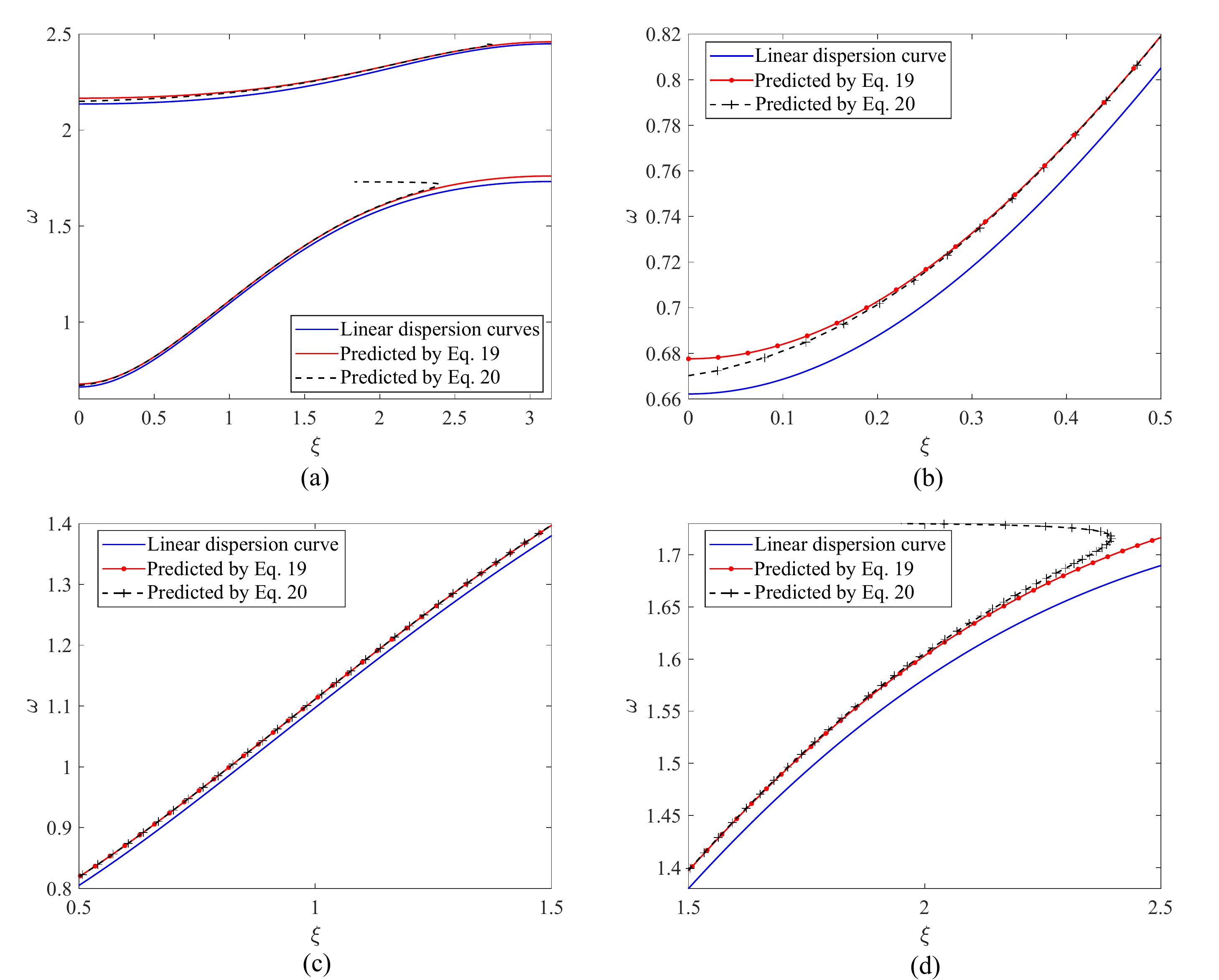}
	\caption{Comparison of dispersion relation shifts estimated by Eq.~\ref{modified_soln1_O1} and Eq.~\ref{modified_soln2_O1}. (a) Corrections of branches. (b), (c), and (d) are zoomed details on in the low-, middle-, and high-wavenumber ranges for the first branch.}
	\label{Modified_Dispersion_Comparison}
\end{figure*}

From Eq.~\ref{modified_soln1_O1} and Eq.~\ref{modified_soln2_O1}, it is clear that that a shift can be induced either in the frequency domain or in the wavenumber domain, depending on whether initial conditions or boundary conditions are imposed. Moreover, the shifts are proportional to $A^2_0$, which means that the dispersion relations are amplitude-dependent. While the two formulations appear interchangeable, the manifestations of the two shifts and their effects on the dispersion relation tuning are profoundly different. In Fig.~\ref{Modified_Dispersion_Comparison}, we plot the corrected dispersion branches predicted by our analysis with the following parameters: $m_1=1$, $m_2=0.5$, $K_2=J_2=G_2=G_4=1$, $\epsilon_1=0.1$, $A_0=1$. The curves calculated from Eq.~\ref{modified_soln1_O1} and Eq.~\ref{modified_soln2_O1} coincide over a large range of wavenumbers, but significant deviations are observed when $\xi$ approaches $0$ or $\pi$. In the latter limit, the dispersion relation predicted by Eq.~\ref{modified_soln2_O1} diverges and loses physical meaning (a similar issue is reported in \cite{CHAKRABORTY_2001} where perturbation expansion is applied for the wavenumber), while Eq.~\ref{modified_soln1_O1} shows a robust estimation on the frequency shift over the whole Brillouin zone. 

\begin{figure} [!htb]
	\centering
	\includegraphics[scale=0.6]{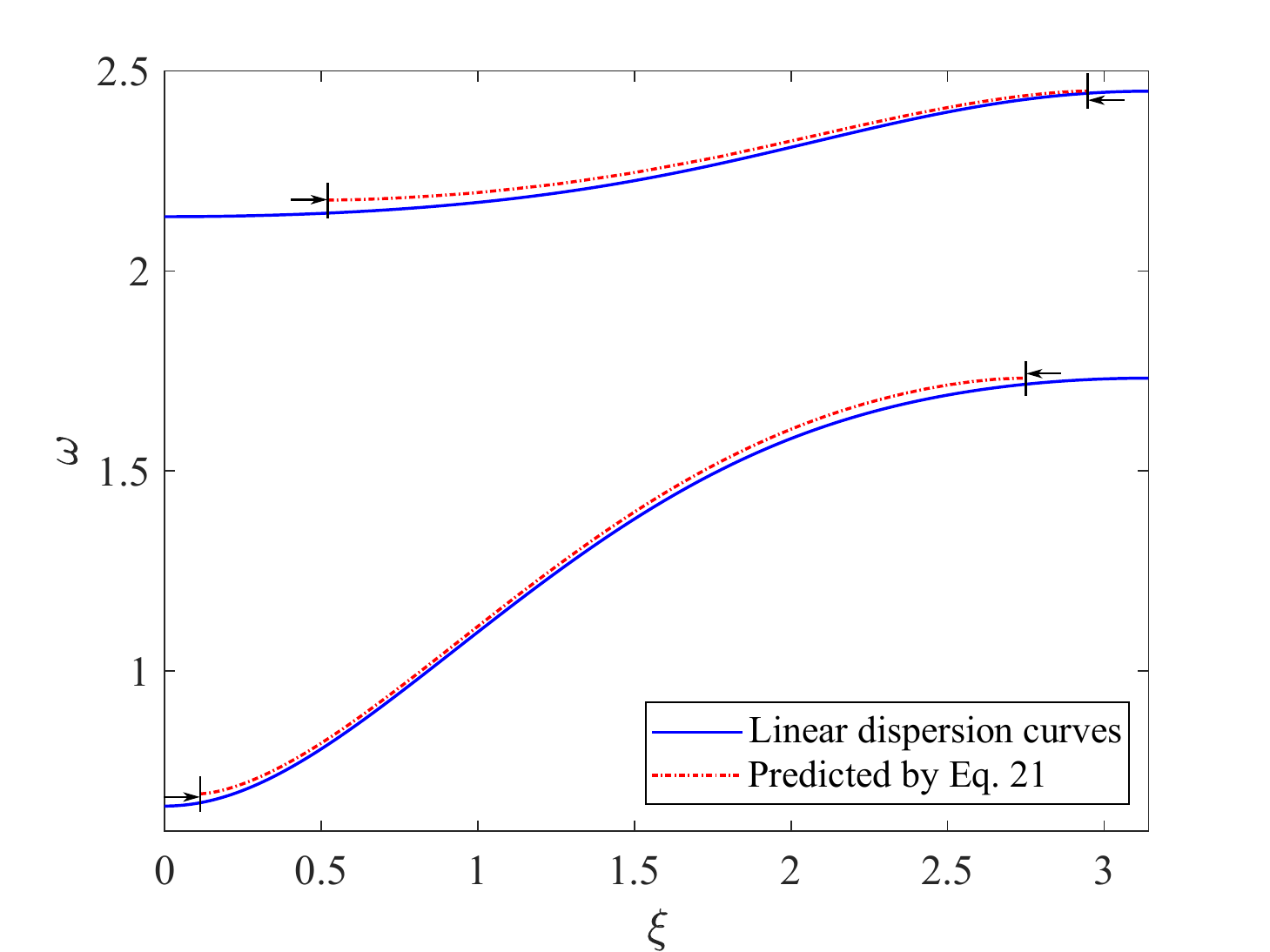}
	\caption{Dispersion shift predicted by Eq.~\ref{secularity_eqn_new}.}
	\label{Dispersion_relation_new}
\end{figure}

To resolve this discrepancy, we have to realize that the Bloch condition (Eq.~\ref{soln2_O1}), which is written in terms of wavenumber, needs to be updated once a wavenumber shift is determined, which means that the original wavenumber $\xi$ should be replaced by the modified one $\xi-\epsilon_1 \beta/s$ (this expression can be obtained by plugging Eq.~\ref{Amplitude} in Eq.~\ref{soln1_O1} and combining the phase terms). The updated Bloch condition leads to a new equation (replacing Eq.~\ref{secularity_eqn}) for the elimination of the secular terms at $O(\epsilon_1)$:
\begin{equation}\label{secularity_eqn_new}
\frac{\partial A}{\partial \tau}+\lambda_0 \sin \left( \xi-\epsilon_1\frac{\beta}{s}\right) \frac{\partial A}{\partial s}+i\mu \abs{A}^2 A=0
\end{equation}
where $\lambda_0=\frac{K_2 \abs{\phi_u}^2 }{\omega \bar{m}}$. 
Substituting Eq.~\ref{Amplitude} in Fig.~\ref{secularity_eqn_new} leads to the following equations for $\alpha$ and $\beta$
\begin{equation}\label{secularity_eqn_new2}
\begin{split}
\frac{\partial \alpha}{\partial \tau}+\lambda_0 \sin\left( \xi-\epsilon_1\frac{\beta}{s}\right) \frac{\partial \alpha}{\partial s}&=0
\\
\frac{\partial \beta}{\partial \tau}+\lambda_0 \sin\left( \xi-\epsilon_1\frac{\beta}{s}\right)  \frac{\partial \beta}{\partial s}&=\mu \alpha^2
\end{split}
\end{equation}
If the wavenumber shift $\epsilon_1\beta/s$ is small when compared to $\xi$, the above system of equations can be reduced to Eq.~\ref{Amplitude_eqns} to the first order approximation. In all other cases, we need to solve Eq.~\ref{secularity_eqn_new2} to find the solution of $\beta$. The particular solution of $\beta$ can be assumed as $\beta^*=C_1 s=C_1\epsilon_1 n$ if plane wave solution is allowed, where $C_1$ is a constant. Substituting it into Eq.~\ref{secularity_eqn_new2} and noticing that  $\alpha=A_0$ under boundary excitation with constant amplitude $A_0$, yields the transcendental equation for $C_1$
\begin{equation}\label{secularity_eqn_new3}
\lambda_0 C_1 \sin \tilde{\xi}=\mu A_0^2
\end{equation}
where $\tilde{\xi}=\xi- \epsilon_1 C_1$. Once the root of Eq.~\ref{secularity_eqn_new3} has been numerically determined, the corrected dispersion relation can be computed using the updated wavenumber $\tilde{\xi}$, resulting in the curves of Fig.~\ref{Dispersion_relation_new}. It can be noticed that the spurious dispersive behavior observed in Eq.~\ref{Modified_Dispersion_Comparison} (d) for large values of $\xi$ is successfully removed.

This result leads to a few interesting observations.  When a correction in $\xi$ is enforced (as a result of an imposed boundary condition), the corrected dispersion curves are not defined over the entire reduced Brillouin zone. This is in contrast with the frequency correction \cite{Narisetti_2010}, which is defined for every $\xi$. This implies that, at certain wavenumbers (especially, approaching the $0$ and $\pi$ limits), wave propagation is forbidden in the nonlinear system if a boundary condition is imposed. The reasonableness and physical implications of this result will be substantiated via numerical simulations in the following chapters. 
In conclusion, regardless of whether boundary conditions or initial conditions are considered, cubic nonlinearity results in a tangible modification of the dispersion relation. 
Since imposing boundary conditions is vastly more practical than prescribing initial conditions in most problems of engineering relevance, we focus on this case in the remainder of the paper. Then, the fundamental solution with the updated wavenumber shift $\tilde{\xi}$ can be expressed compactly as 
\begin{equation}\label{modified_soln2_O1_new}
\mathbf{u}_n^0=A_0 \bm{\phi} e^{i\left(  \tilde{\xi} n-\omega t\right)  }+c.c.
\end{equation}.

\subsection{Second harmonic generation and its correction shift}
It has been shown that, in the presence of quadratic nonlinearity, the complete solution at $O(\epsilon_2)$ encompasses a non-oscillatory term and an oscillatory contribution, the latter comprising the solution of the corresponding homogeneous problem and the particular (forced) solution due to the forcing term $\mathbf{f}^2$ \cite{Tournat_2013}. The question we address here is whether the presence of cubic nonlinear springs in the foundation introduces additional tuning effects that are appreciated at the level of these second harmonic contributions. To this end, in the following we derive a general expression for the second harmonic solution that explicitly incorporates any spectral modulation due to such nonlinear interplay.

If we substitute the associated fundamental solution (i.e., Eq.~\ref{modified_soln2_O1_new}) in Eq.~\ref{order2}, the expression for the forcing function (note that only the first component $f_u^2$ of $\mathbf{f}^2$ is non-zero) can be determined as (details about the derivation are given in Appendix)
\begin{align}\label{forcing_term2}
\begin{split}
f_u^2=-8iK_3A^2_0 \phi_u^2\sin \tilde{\xi} \sin^2 \frac{\tilde{\xi}}{2} e^{2i\tilde{\theta}_n}+c.c.
\end{split}
\end{align}
where $\tilde{\theta}_n=\tilde{\xi} n-\omega t$ and $\tilde{\xi}$ is the corrected wavenumber introduced above and, by working with $\tilde{\xi}$, we implicitly assume that the excitation is prescribed as a boundary condition. Since the coefficient in front of $e^{2i\tilde{\theta}_n}$ is purely imaginary, this expression suggests a phase shift in the second harmonic.

\subsubsection{Forced solution and its correction}

Following the standard procedure for heterogeneous PDEs, the forced solution can be found as
\begin{equation}\label{forced_soln_O2}
\mathbf{u}_n^{2f}=B_1\bm{\varphi}e^{2i\tilde{\theta}_n}+c.c.
\end{equation}
where the coefficient can be obtained by plugging Eq.~\ref{forced_soln_O2} in Eq.~\ref{order2}, yielding
\begin{equation}\label{forced_soln_coe}
B_1\bm{\varphi}=\left[-4\omega^2 \mathbf{M} + \mathbf{K}(\tilde{\xi})\right]^{-1} \mathbf{\bar{f}}^2
\end{equation}
where $\mathbf{\bar{f}}^2=\begin{Bmatrix}-8iK_3A^2_0 \phi_u^2\sin \tilde{\xi} \sin^2 \frac{\tilde{\xi}}{2} \\ 0\ \end{Bmatrix}$ and $\bm{\varphi}$ is subjected to the same normalization followed for the fundamental modal vector $\bm{\phi}$. The phase $2i\tilde{\theta}_n$ indicates that $\mathbf{u}_n^{2f}$ represents a harmonic wave traveling with $2\tilde{\xi}$ and $2\omega$, i.e., twice the frequency and twice the (corrected) wavenumber of the fundamental harmonic $\mathbf{u}_n^0$. While the doubling of frequency and wavenumber is an expected feature of the forced response in systems with quadratic nonlinearity \cite{Tournat_2013, Mehrem_2017}, the dependence on $\tilde{\xi}$ introduces a new spectral control that is germane to the problem with additional cubic nonlinearity. In essence, cubic nonlinearity introduces an additional wavenumber shift in the signature of the quadratically-generated forced second harmonic. We conclude that the interplay between the two types of nonlinearity results in a new amplitude-dependent correction mechanism that endows traveling waves with an additional self-manipulation capability. We now proceed to check whether a similar effect is observed in the homogeneous part of the SHG. 

\subsubsection{Homogeneous solution and its correction}

The homogeneous solution can be expressed as 
\begin{equation}\label{homo_soln_O2}
\mathbf{u}_n^{2h}=iB_2\bm{\chi}e^{i\psi_n}+c.c.
\end{equation}
In contrast with the forced response, here the phase $\psi=\xi(2\omega)n-2\omega t$ involves the frequency-wavenumber pair predicted by the branch of the band diagram available at $2\omega$, thus conforming to the dispersion relation of the corresponding liner problem. Accordingly, $\bm{\chi}=\begin{Bmatrix} \chi_u \\ \chi_v\\ \end{Bmatrix}$ is the modal vector at $\left( \xi(2\omega),2\omega\right)$, and the imaginary constant $i$ is assume a priori in Eq.~\ref{homo_soln_O2} is to simplify the analysis.

In the spirit of multiple scale analysis, the amplitude term $B_2$ is, in general, taken to be dependent upon the slow variables $s$ and $\tau$, and can be expressed as 
\begin{equation}\label{Amplitude_SHG_homo}
B_2\left(s,\tau \right) = a\left(s,\tau \right)e^{-ib\left(s,\tau \right)}
\end{equation}
The full characterization of the homogeneous response relies on the non-trivial task of determining $B_2\left(s,\tau \right)$.
To this end, we introduce a manipulation step aimed at virtually extending to the second harmonic the treatment, classically reserved for fundamental harmonic, that we followed in section 1.1 for the determination of $A(s, \tau)$. Specifically, we formally combine the homogeneous second harmonic and the fundamental harmonic into a new baseline expression, which we will treat as the ``new fundamental solution" at $O(1)$:
\begin{equation}\label{revisit_O1}
\hat{\mathbf{u}}_n^0=A\bm{\phi}e^{i\theta_n}+A^*\bm{\phi}^*e^{-i\theta_n}+i\epsilon_2 B_2\bm{\chi}e^{i\psi_n}-i\epsilon_2 B^*_2\bm{\chi}^*e^{-i\psi_n}
\end{equation}
Accordingly, the equation at $O(\epsilon_1)$ is also updated, and the cubic nonlinear term in $\mathbf{f}^1$ becomes
\begin{widetext}
\begin{align}\label{forcing_term1}
\begin{split}
f^1_{cubic}&= -G_4 \left( v^0_n\right) ^3 = -G_4 \left(A\phi_v e^{i\theta_n}+A^*\phi_v^*e^{-i\theta_n}+i\epsilon_2 B_2\chi_v e^{i\psi_n}-i\epsilon_2 B^*_2\chi_v^*e^{-i\psi_n} \right)^3 \\
&= -3G_4\left(  \abs{A}^2 \abs{\phi_v}^2 A \phi_v + 2\epsilon^2_2 \abs{B_2}^2 \abs{\chi_v}^2 A \phi_v \right) e^{i\theta_n}  
-3i\epsilon_2 G_4\left(  2\abs{A}^2 \abs{\phi_v}^2 B_2\chi_v + \epsilon^2_2\abs{B_2}^2 \abs{\chi_v}^2 B_2\chi_v \right) e^{i\psi_n} +c.c.
\end{split}
\end{align}
\end{widetext}
where higher order terms have been omitted. 
Following the same logic described in section $1.1$, the removal of the secular terms for the wave modes associated with $e^{i\theta_n}$ and $e^{i\psi_n}$ requires that
\begin{multline}\label{secular_1}
\frac{\partial A}{\partial \tau}+\lambda_0 \sin\left( \xi-\epsilon_1\frac{\beta}{s}\right) \frac{\partial A}{\partial s}\\+i\left( \mu \abs{A}^2 +\epsilon^2_2\eta\abs{B_2}^2 \right) A=0
\end{multline}
and
\begin{multline}\label{secular_2}
\frac{\partial B_2}{\partial \tau}+\lambda'_0 \sin\left( \xi(2\omega)-\epsilon_1\frac{b}{s}\right) \frac{\partial B_2}{\partial s}\\
+i\left( \mu' \abs{A}^2 +\epsilon^2_2\eta' \abs{B_2}^2 \right) B_2=0
\end{multline}
where $\eta = \frac{3G_4 \abs{\chi_v}^2 \abs{\phi_v}^2}{\omega\bar{m}}$; $\lambda'_0=\frac{K_2 \abs{\chi_u}^2 }{ 2\omega  \bar{m}}$; $\mu'=\frac{3G_4 \abs{\phi_v}^2\abs{\chi_v}^2}{ 2\omega  \bar{m}}$; $\eta' = \frac{3G_4 \abs{\chi_v}^4}{4\omega \bar{m}}$.
Ignoring the high order term $O(\epsilon^2_2)$, Eq.~\ref{secular_1} reduces back to Eq.~\ref{secularity_eqn_new}, whose solution yields the dispersion shift in the fundamental harmonic, as discussed before. Similarly, neglecting high order term in Eq.~\ref{secular_2}, the equation becomes
\begin{equation}\label{secular_2_new}
\frac{\partial B_2}{\partial \tau}+\lambda'_0 \sin\left( \xi(2\omega)-\epsilon_1\frac{b}{s}\right)  \frac{\partial B_2}{\partial s}+i \mu' \abs{A}^2 B_2=0
\end{equation}
which is a partial differential equation in $s$ and $\tau$ controlling the slow spatio-temporal evolution of $B_2$. Substituting Eq.~\ref{Amplitude_SHG_homo} in Eq.~\ref{secular_2_new} and assuming $b=C_2 s=\epsilon_1 C_2 n$ (an apriori assumption which leads to a traveling wave solution if $C_2$ can be solved as a real number), yields a transcendental equation formally similar to Eq.~\ref{secularity_eqn_new3}, which can be solved numerically for $C_2$. With $C_2$, the homogeneous solution can be rewritten as
\begin{equation}\label{homo_soln_O2z-new}
\mathbf{u}_n^{2h}=ia\bm{\chi}e^{i\hat{\psi}_n}+c.c.
\end{equation}
where $\hat{\psi}_n= \hat{\xi}n-2\omega t$ and $\hat{\xi}=\xi(2\omega)-\epsilon_1 C_2$. This result shows that the Q-C interplay leads to a new dispersion shift that is germane to the homogeneous second harmonic.

\subsubsection{Phase matching conditions}

A careful inspection on Eq.~\ref{forced_soln_coe} suggests that the above procedures for finding the solutions at $O(\epsilon_2)$ works only when $-4\omega^2 \mathbf{M} + \mathbf{K}\equiv \mathbf{Q}$ is invertible. It is possible, however, that $\mathbf{Q}$ is singular ($det(\mathbf{Q})=0$). If that is the case, the term containing the variable $2\tilde{\theta}$ in the forcing function $\mathbf{f}^2$ satisfies the linear eigenvalue problem of Eq.~\ref{eigenvaule_prob}. In other words, the forcing function resonates with the homogeneous solution (i.e., $2\tilde{\theta}_n=\psi_n$) and the solution becomes secular. Such scenario is referred to as phase matching condition (PMC) and in general has been shown to be conducive to stronger nonlinear response signatures \cite{JIAO2018, Jiao_2018_PRA}. The solution, which is assumed to grow in direct proportion to the propagation distance, is expressed as
\begin{equation}\label{PMC_soln_O2}
\mathbf{u}_n^{2}=D\bm{\chi}ne^{2i\tilde{\theta}_n}+c.c.
\end{equation}
Substituting it in Eq.~\ref{order2}, yields
\begin{equation}\label{PMC_eqn_O2}
D\left(-4\omega^2 \mathbf{M} + \mathbf{K(\tilde{\xi})} \right) \bm{\chi}n+D\mathbf{p}=\mathbf{\bar{f}}^2
\end{equation}
where $\mathbf{p}=-K_2\begin{Bmatrix} 2i\chi_u \sin 2\tilde{\xi} \\ 0\\ \end{Bmatrix}$. The first term in the LHS of Eq.~\ref{PMC_eqn_O2} automatically vanishes since it satisfies the eigenvalue problem Eq.~\ref{eigenvaule_prob}, as stated above. Then, the coefficient $D$ can be determined as: $D=\frac{4K_3A^2_0\phi^2_u \sin \tilde{\xi} \sin^2 \frac{\tilde{\xi}}{2}} {K_2\chi_u \sin 2\tilde{\xi}}$.

As we mentioned earlier in the manuscript, this type of nonlinear system also supports a non-oscillatory solution - a general nonlinear effect in anharmonic lattices \cite{Mehrem_2017}. For completeness, let us briefly re-examine the forcing term at $O(\epsilon_2)$ (see Appendix), in which we notice that a constant force can be induced in each quadratic spring by the underlined terms in Eq.~\ref{A1}. 
\begin{align}\label{non_osci_soln}
\begin{split}
F_{s}&=2 \epsilon_2 K_3A^2_0 \abs{\phi_u}^2 \left(1-e^{-i\tilde{\xi}} \right)\left(1-e^{i\tilde{\xi}} \right) \\
&=4\epsilon_2 K_3A^2_0 \abs{\phi_u}^2(1-\cos \tilde{\xi})
\end{split}
\end{align}
This constant force results in a stationary displacement $\mathbf{u}^{2s}_n$ in the system, and such static problem can be easily solved once the size (i.e., the total number of degrees of freedom) of the system is specified. In conclusion, the complete solution at $O(\epsilon_2)$ can be expressed as \\
If $2\tilde{\theta}_n\neq\psi_n $ (Non-PMC):
\begin{align}
\begin{split}\label{non_PMC_soln_complete}
\mathbf{u}^2_n=\mathbf{u}^{2s}_n+B_1\bm{\varphi}e^{2i\tilde{\theta}_n}+ia\bm{\chi}e^{i\hat{\psi}_n}+c.c. 
\end{split}
\end{align}
If $2\tilde{\theta}_n = \psi_n $ (PMC):
\begin{align}
\begin{split}\label{PMC_soln_complete}
\mathbf{u}^2_n=\mathbf{u}^{2s}_n+D\bm{\chi}ne^{2i\tilde{\theta}_n}+c.c.
\end{split}
\end{align}
Imposing the absence of second harmonic and of non-oscillatory terms at the boundary ($n=0$), it yields $b_0=0$ and $a=iB_1 \varphi_v/\chi_v$, and the unknowns in $\mathbf{u}^{2s}_n$ can be fully determined. 

\begin{figure*} [!htb]
	\centering
	\includegraphics[scale=0.55]{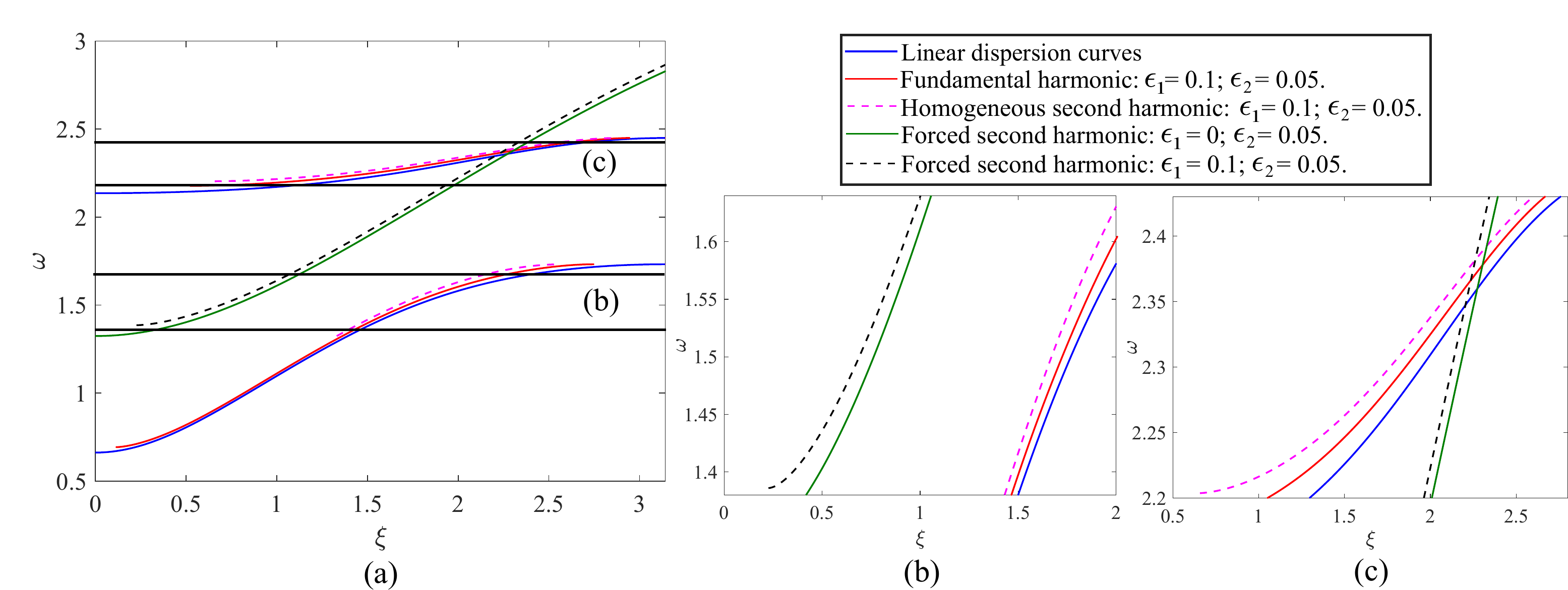}
	\caption{(a) Dispersion shifts in the fundamental and second harmonics. (b-c) Zoomed details of the (b) lower branch and (c) upper branch, with amplitude $A_0=1$.}
	\label{Modified_Dispersion_Comparison_Complete}
\end{figure*}

Equations~\ref{non_PMC_soln_complete} and~\ref{PMC_soln_complete} shed light over some unique characteristics of the nonlinearly-generated second harmonics, which are summarized below: \\
1) The shift in wavenumber introduced by the cubic nonlinearity at $O(\epsilon_1)$, is also observed in the forced solution at the second harmonic. However, the shift at the second harmonic is magnified by a factor of $2$ (i.e., 2$\Delta \xi =2(\tilde{\xi}-\xi)$), which implies that the spectral signatures of the forced second harmonic can be modified to a larger extent than the fundamental harmonic by playing with the amplitude of excitation. \\
2) We have shown that the interaction between the quadratic and cubic nonlinearities produces a new dispersion shift in the second harmonics, which is conceptually distinct from the shift in the fundamental harmonic. In other words, at a given frequency, two shifts are defined: the one experienced by a fundamental harmonic excited directly at that frequency and the one experienced by the second harmonic generated by an excitation with half frequency content. We note that the dispersion shift in the homogeneous second harmonic (dashed red lines) is significantly larger (precisely twice larger) when compared to the shift in the fundamental solution (red lines) at that frequency. A visual comparison of these shifts is provided in Fig.~\ref{Modified_Dispersion_Comparison_Complete} for different frequencies and strengths of nonlinearity. This unique feature enables an additional array of wave manipulation opportunities, as will be discussed in the application section below.\\
3) When PMC is achieved, the amplitude of the second harmonic grows with the propagation distance, and the full nonlinear response  conforms to the linear dispersion relation. 
As discussed in \cite{JIAO2018, Jiao_2018_PRA}, this unique property has important practical implications. The energy tunneling induced by the SHG from the fundamental to second harmonic can be boosted by enabling PMC. \\
4) In the presented multiple scales analysis, the selection of the scale variables is deliberately chosen to capture the main effects of quadratic and cubic nonlinearities in the system. A more general choice, up to $O(\epsilon_2)$ would include: $\theta_n=\xi n-\omega t$; $s_1=\epsilon_1 n$; $\tau_1=\epsilon_1 t$; $s_2=\epsilon_2 n$; $\tau_2=\epsilon_2 t$. However, the two additional variables $s_2$ and $\tau_2$ will not add any new information to the existing solutions in our context. This can be explained by recalling that the main purpose of introducing the slow variables $s$ and $\tau$ is to capture the shift in the dispersion relation. To the first order approximation (i.e., at $O(\epsilon_1)$), this is brought about exclusively by the cubic nonlinearity, while the shift due to quadratic nonlinearity is negligible \cite{Ganesh_2013}. 
Another consequence of this argument is that the presented perturbation analysis also works when $\epsilon_2$ is comparable to $\epsilon_1$, i.e., $\epsilon_1\approx \epsilon_2$. Under these circumstances, Eq.~\ref{order2} and Eq.~\ref{order3} can be combined into one equation with the two forcing functions $\mathbf{f}^1$ and $\mathbf{f}^2$, and the rest of the procedure remains the same.


\section{Full-scale simulations}

In this section, we perform full-scale numerical simulations to validate our analytical model and support the findings detailed in section 2. The wave response of a finite chain is obtained by integrating the system of nonlinear equations (Eq.~\ref{governing_eqns}) using the Verlet Algorithm \cite{Swope_1982}. In order to obtain clean signatures of the desired harmonics, it is imperative to work with traveling waves that are as monochromatic as possible. This can be achieved by prescribing sustained harmonic excitations at the boundary. However, this route is impractical with finite-size systems as reflections from the boundaries would soon result in standing waves (i.e., vibration states). 

To prescribe purely harmonic traveling waves while avoiding boundary reflections, we implement an absorbing layer of $30$ unit cells endowed with viscous damping at the end of the main chain which consists of $40$ undamped unit cells. The coefficient profile for the $n^{th}$ unit cell in the absorbing layer is given by 
\begin{equation}\label{PML_profile}
\mathbf{C}_n=C_0\left(\frac{n}{N_{\mathrm{PML}}} \right) ^ \gamma \mathbf{I}
\end{equation}
where $C_0=3$ is the maximum damping coefficient, $N_{\mathrm{PML}}=30$ denotes the number of unit cells in the layer, $\gamma$ is set to $3$ in the simulation, and $\mathbf{I}$ is the $2\times2$ identity matrix. A continuous harmonic excitation is applied to mass $m_2$ in the first unit cell of the chain. 
The spatio-temporal response computed at each cell site over the considered time interval is transformed to the Fourier domain via two-dimensional Discrete Fourier Transform (2D-DFT). The spectral components $(\xi, \omega)$ of the response, to be compared against the band diagrams from unit cell analysis, are extracted by tracking the maximum values of the 2D-DFT data in different frequency intervals.

\subsection{Numerical estimation of dispersion shifts in the fundamental harmonic}
\begin{figure*} [!htb]
	\centering
	\includegraphics[scale=0.6]{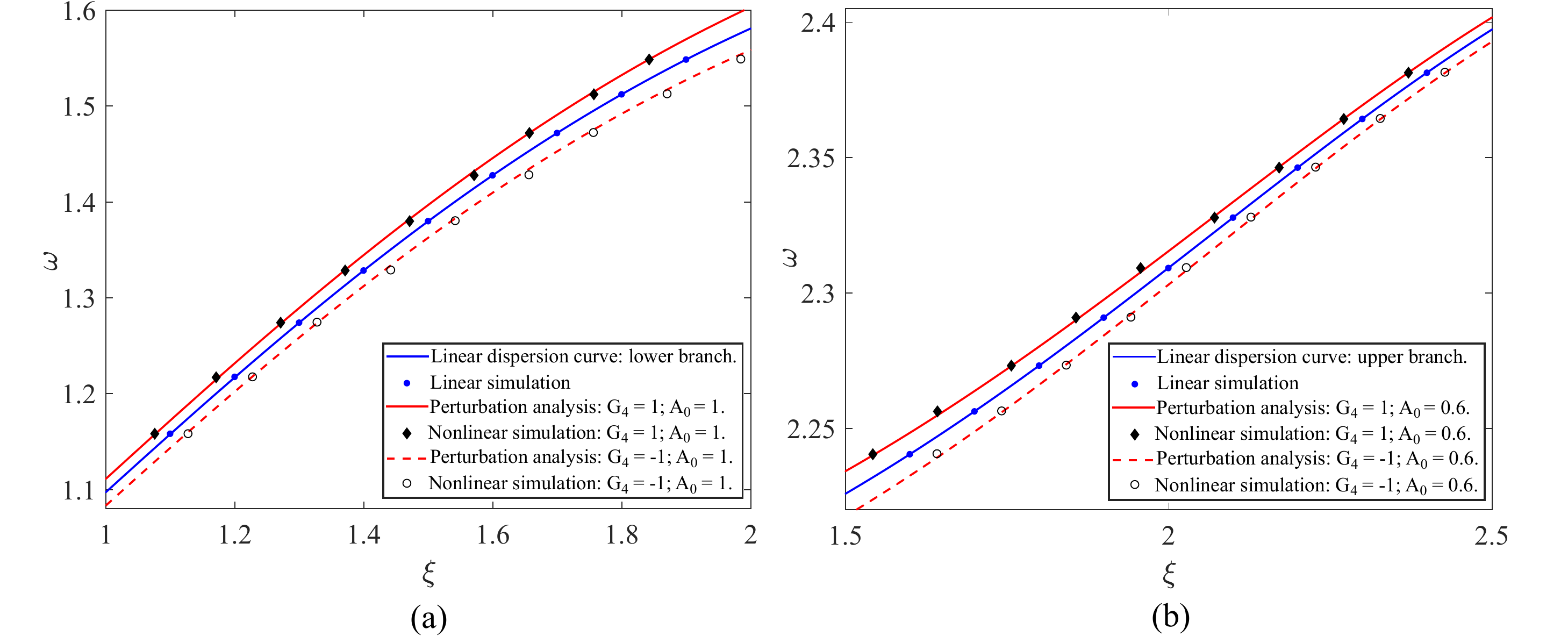}
	\caption{Numerical estimation of the dispersion shift in the lower (a) and upper (b) branches of the band diagram, compared with the analytical model based on perturbation analysis.}
	\label{Modified_Dispersion_simulations}
\end{figure*}

The system parameters are $m_1=1$, $m_2=0.5$, $K_2=J_2=G_2=1$, $\epsilon_1=0.1$. The cubic nonlinearity can be positive ($G_4=1$) or negative ($G_4=-1$), which corresponds to a hardening or softening nonlinear foundation, respectively. The amplitude of the excitation $A_0$ is set to $1$ for the lower branch correction (Fig.~\ref{Modified_Dispersion_simulations}(a)) and to $0.6$ for the upper branch correction (Fig.~\ref{Modified_Dispersion_simulations}(b)).

Fig.~\ref{Modified_Dispersion_simulations} shows that the numerical results match the multiple scales analysis predictions. The branches are shifted upward or downward depending on whether hardening or softening is considered. An interesting observation can be made about the regime close to the cutoff frequencies, where the result obtained by Eq.~\ref{modified_soln2_O1} (Fig.~\ref{Modified_Dispersion_Comparison}) diverged spuriously. In these regimes, the responses from simulations tend to be unstable and monochromatic conditions cannot be established effectively. This can be qualitatively explained by invoking the fact that the dispersion curves become flat around the cutoff frequencies, resulting in large wavenumber shifts (if boundary conditions are imposed) that violate the weak nonlinearity assumption. As a result, the dispersion shift predicted by Eq.~\ref{modified_soln2_O1} can be used as a guideline to check validity of the assumptions underpinning our perturbation analysis for a given system and excitation conditions.

\subsection{Numerical estimation of dispersion shift in SHG}

\begin{figure*} [!htb]
	\centering
	\includegraphics[scale=0.6]{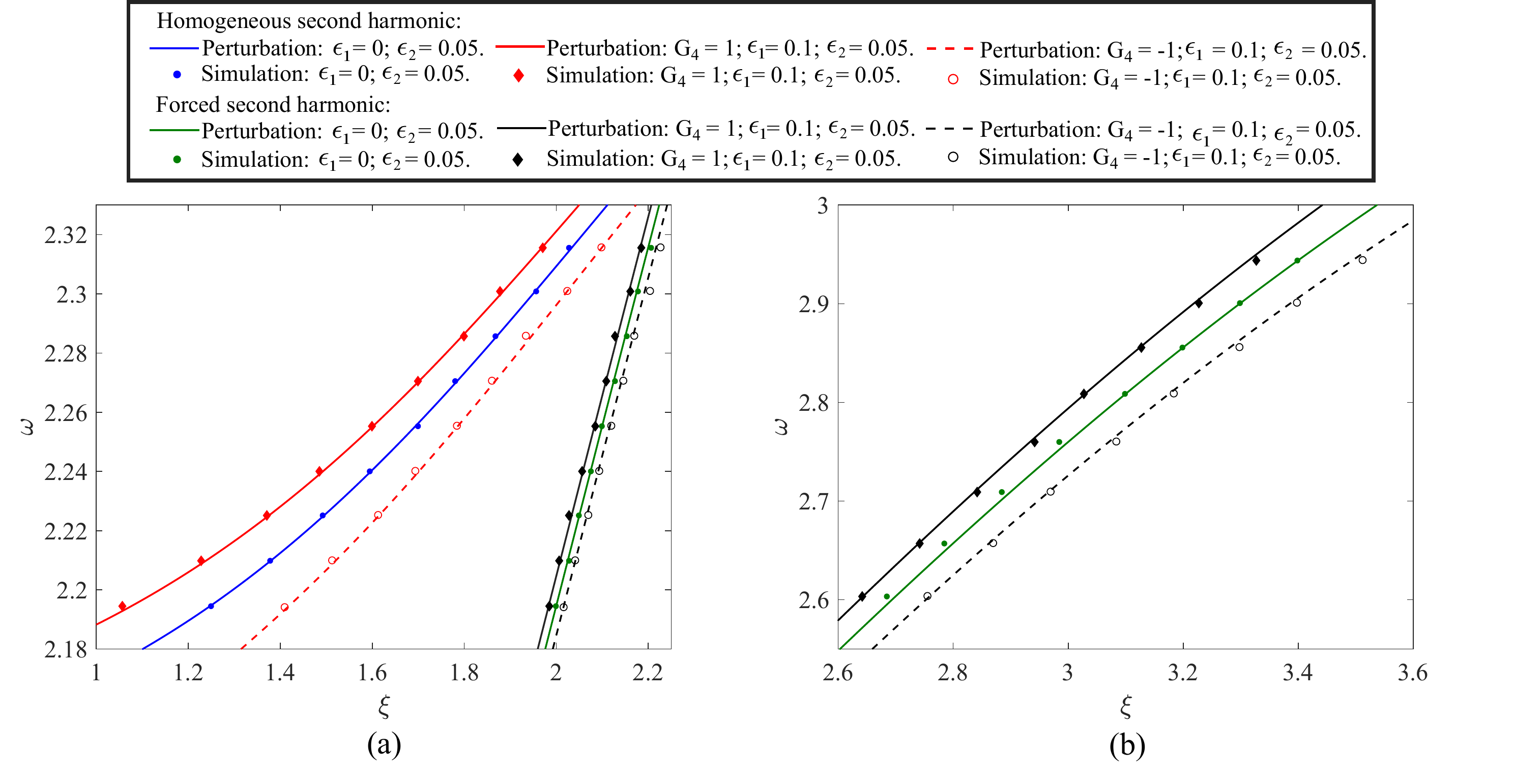}
	\caption{Numerical estimation of dispersion shift in the second harmonics (a) in frequency range where homogeneous and forced solutions co-exist ($A_0=0.6$), and (b) in frequency range where only the forced solution exists ($A_0=1$).}
	\label{Dispersion_relation_Simulations_SHG}
\end{figure*}

So far, we have numerically matched only the shifts described by Eq.~\ref{modified_soln2_O1_new}, which only affect the fundamental harmonic. Now, we incorporate quadratic nonlinearity ($K_3=1$ and $\epsilon_2=0.05$) in the springs of the main chain to elicit second harmonic generation and estimate its spectral shift. 
The spectral content of the forced second harmonic is determined by the $(\tilde{\xi},2\omega)$ pair, and the homogeneous one is dictated by the $\hat{\xi}- 2\omega$ curve. Both wavenumbers $\tilde{\xi}$ and $\hat{\xi}$ are subjected to (different) correction shifts and depend on the excitation amplitude $A_0$. Fig.~\ref{Dispersion_relation_Simulations_SHG} shows the numerical results of the dispersion shift in the second harmonics. Two scenarios can be identified. In a certain frequency range, the forced and homogeneous solutions co-exist, as shown in Fig.~\ref{Dispersion_relation_Simulations_SHG}(a). In the frequency range where no modes are available, only the forced solutions is observed, as shown in Fig.~\ref{Dispersion_relation_Simulations_SHG}(b). In both cases, the agreement between the analytical and numerical results is remarkable. Comparing Fig.~\ref{Modified_Dispersion_simulations}(a) against Fig.~\ref{Dispersion_relation_Simulations_SHG}(b), and Fig.~\ref{Modified_Dispersion_simulations}(b) against Fig.~\ref{Dispersion_relation_Simulations_SHG}(a), we can easily conclude that the wavenumber shifts in the homogeneous and forced second harmonics are significantly larger than those exhibited by the fundamental harmonics under the same amplitudes of the excitation, consistently with the prediction of the multiple scales analysis.

\section{Application to waveguides with self-switching capabilities}

In this section we demonstrate that the extreme tunability of the second harmonic enabled by the cubic nonlinearity allows controlling the spectral characteristics of the nonlinear wave response due to the quadratic nonlinearity, and activate new modal functionalities at the second harmonic. 
One interesting application is to design a family of self-adaptive modal switches that can turn on and off some desired functionalities by simply adjusting the amplitude of excitation. Here we illustrate the richness of opportunities resulting from this idea through two illustrative examples. 


\subsection{Application 1: Optical mode switch via homogeneous solution self-tuning}

The first example exploits the possibility to control the manifestation of the homogeneous second harmonic (due to quadratic nonlinearity) through the cubic term. This allows activating/deactivating or tuning certain modal characteristics typical of the higher-frequency optical regime that are nonlinearly activated in the response via SHG. 
 
\begin{figure*} [!htb]
	\centering
	\includegraphics[scale=0.6]{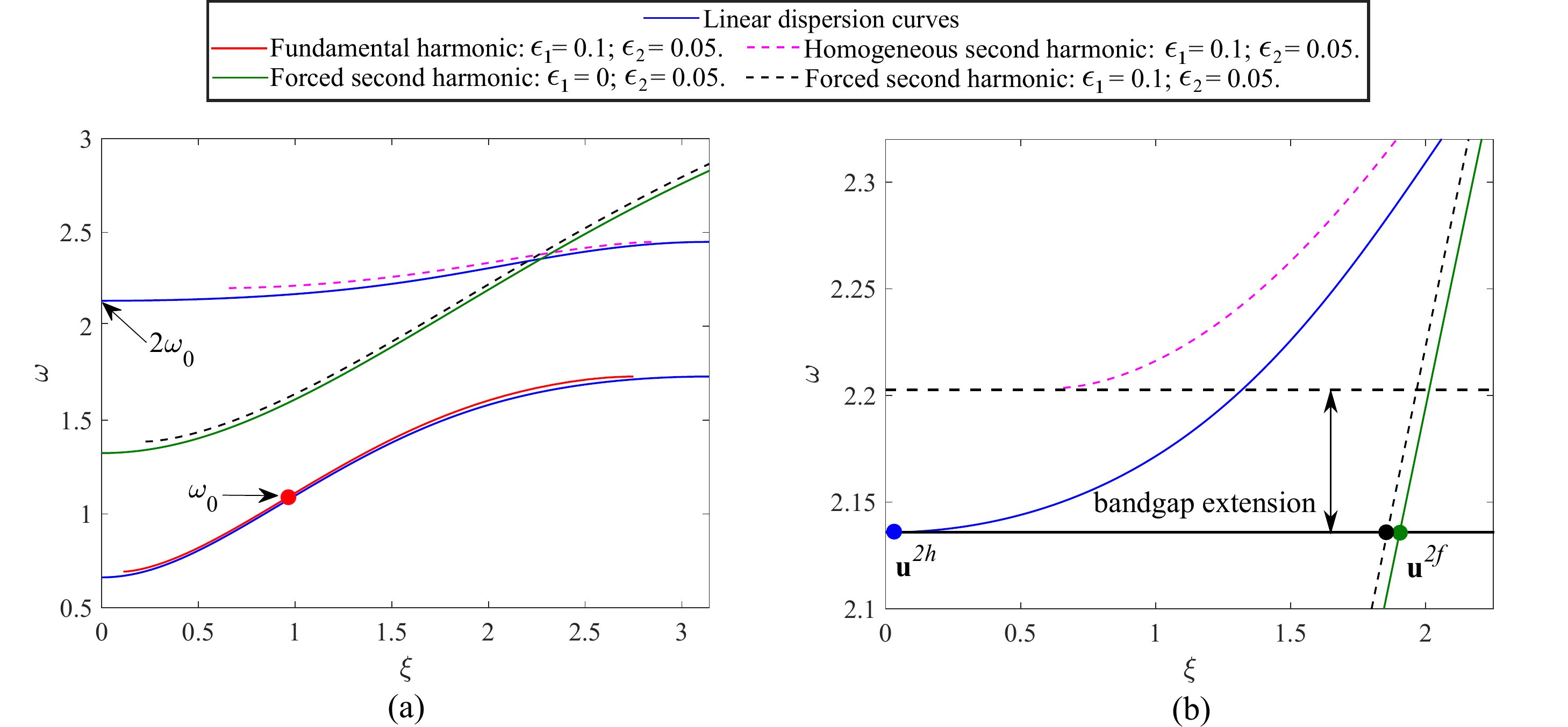}
	\caption{Schematic of a self-adaptive switch which can turn on/off homogeneous second harmonic. (a) Entire band diagram, and (b) zoomed details at the second harmonic frequency.}
	\label{Application1_illustration_all}
\end{figure*}

This application is based on the notion that, when a wave mode (dispersion branch) is available at the second harmonic, the homogeneous solution will display the spectral modal characteristics of that mode \cite{JIAO2018}. If the second harmonic is subjected to a dispersion shift, e.g., through cubic nonlinear correction, the modal characteristics of the activated harmonic can be further manipulated with respect to the default case with only quadratic nonlinearity. This concept is illustrated in Fig.~\ref{Application1_illustration_all} and supported by numerical simulations performed with the same parameters used in section $2$. The excitation frequency $\omega_0$ (red dot in Fig.~\ref{Application1_illustration_all}(a)) is carefully chosen such that the second harmonic $2\omega_0$ is above, but sufficiently close to, the cut-on frequency $\omega^2_{\mathrm{cut-on}}$ of the upper branch. 
Thus, at low amplitude (or without cubic nonlinearity), the homogeneous second harmonic is located at the onset of the upper branch (around the blue dot in Fig.~\ref{Application1_illustration_all}(b)), whose response involves large motion of mass $m_2$ (a typical feature of locally resonant bandgaps). The forced solution is marked by the green dot located on the green line, which represents the locus of the $2\xi-2\omega$ pairs. At high amplitude ($A_0=1$), the dispersion relation is corrected. As a result, the forced second harmonic is shifted from the location of the green dot to the location of the black dot, while the homogeneous second harmonic is in the bandgap extension produced by the upshifted upper branch (magenta dashed line). 

\begin{figure*} [!htb]
	\centering
	\includegraphics[scale=0.4]{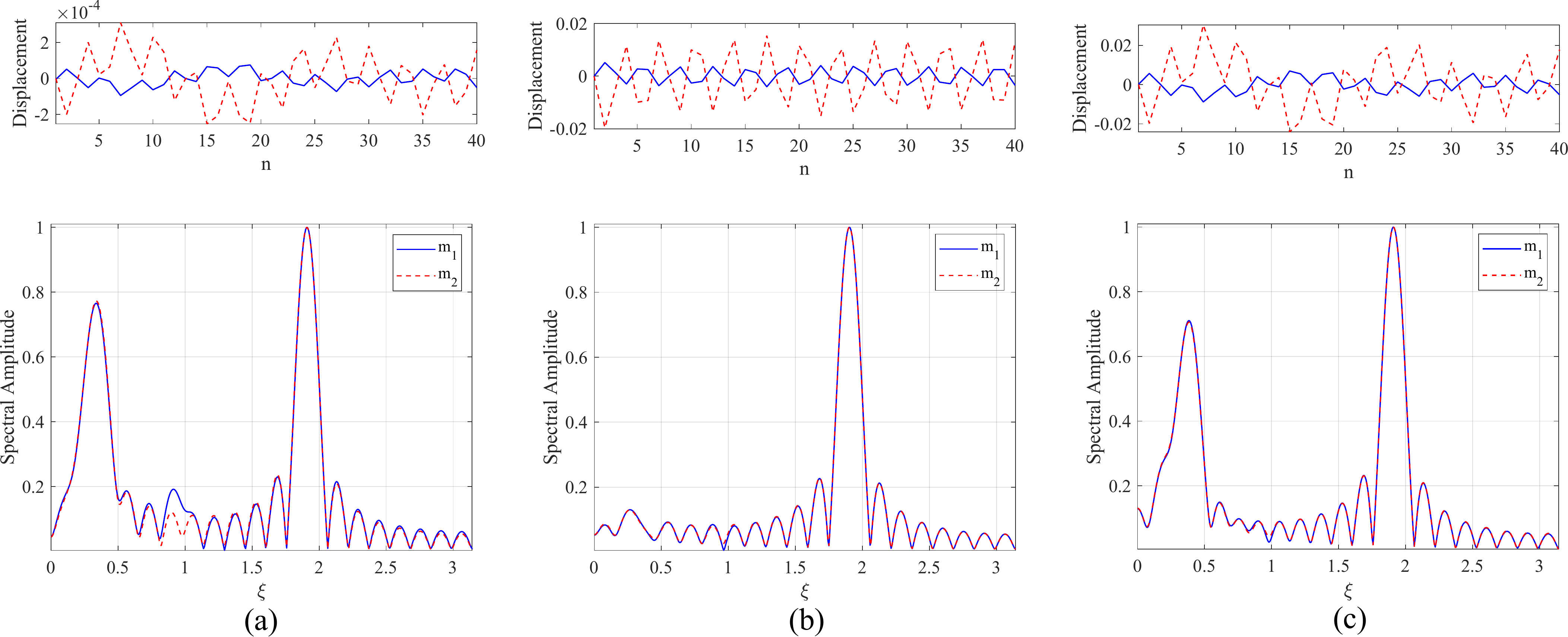}
	\caption{Spatial profile and DFT of the filtered second harmonics. (a) Low amplitude $A_0=0.1$ with $\epsilon_1=0.1$ and $\epsilon_2=0.05$.  (b) High amplitude $A_0=1$ with $\epsilon_1=0.1$ and $\epsilon_2=0.05$. (c) High amplitude $A_0=1$ with  $\epsilon_2=0.05$ (no cubic nonlinearity).}
	\label{Application1}
\end{figure*}

In simulations, the prescribed excitation frequency is chosen to be $\omega_0=1.07$ rad/s such that the second harmonic is around $2\omega_0=2.14$ rad/s, which is close to the cut-on frequency $\omega^2_{\mathrm{cut-on}}$, as desired. A band-pass filter is used to extract the second harmonics from the nonlinear response. Fig.~\ref{Application1} depicts the spatial profiles and the normalized DFT of the filtered second harmonics for mass $m_1$ and mass $m_2$. With a low-amplitude excitation, as shown in Fig.~\ref{Application1}(a), we observe that there are two peaks in the spectral domain: the one with lower wavenumber corresponds to the homogeneous solution, and the other to the forced solution. This is consistent with the spatial profile, where the wave presents an amplitude modulation - a typical feature of waves with two co-existing spectral components. At high amplitude, the first peak is eliminated, and the wavefield becomes monochromatic, as depicted in Fig.~\ref{Application1}(b). In comparison, in Fig.~\ref{Application1}(c) we consider a case where we retain a high amplitude of excitation but we remove the cubic nonlinearity in the system. We see that all the major wave signatures are qualitatively similar to case (a), albeit with higher displacement amplitude. This indicates that the SHG due to quadratic nonlinearity is triggered under large amplitude-excitations, but the additional correction (filtering of the homogeneous component) due to the cubic term in the foundation is lost. In essence, we have demonstrated that, by playing with the amplitude of the excitation and with cubic nonlinearity to exploit, we obtain a double tuning effect on the wave: 1) we trigger and control the strength of the second harmonic (this effect would be observed even without cubic term); 2) we selectively determine whether the homogeneous solution is observed in the second harmonic in addition to the forced one (this effect is germane to the case with double nonlinearity) and therefore we determine whether two distinct wavenumbers and modal characteristics will co-exist in the response. In summary, by controlling the amplitude, not only do we control the strength of the second harmonic; we also inherently dictate its spectral and modal content.

\subsection{Application 2: Adaptive optical mode switch via PMC tuning}
\begin{figure*} [!htb]
	\centering
	\includegraphics[scale=0.7]{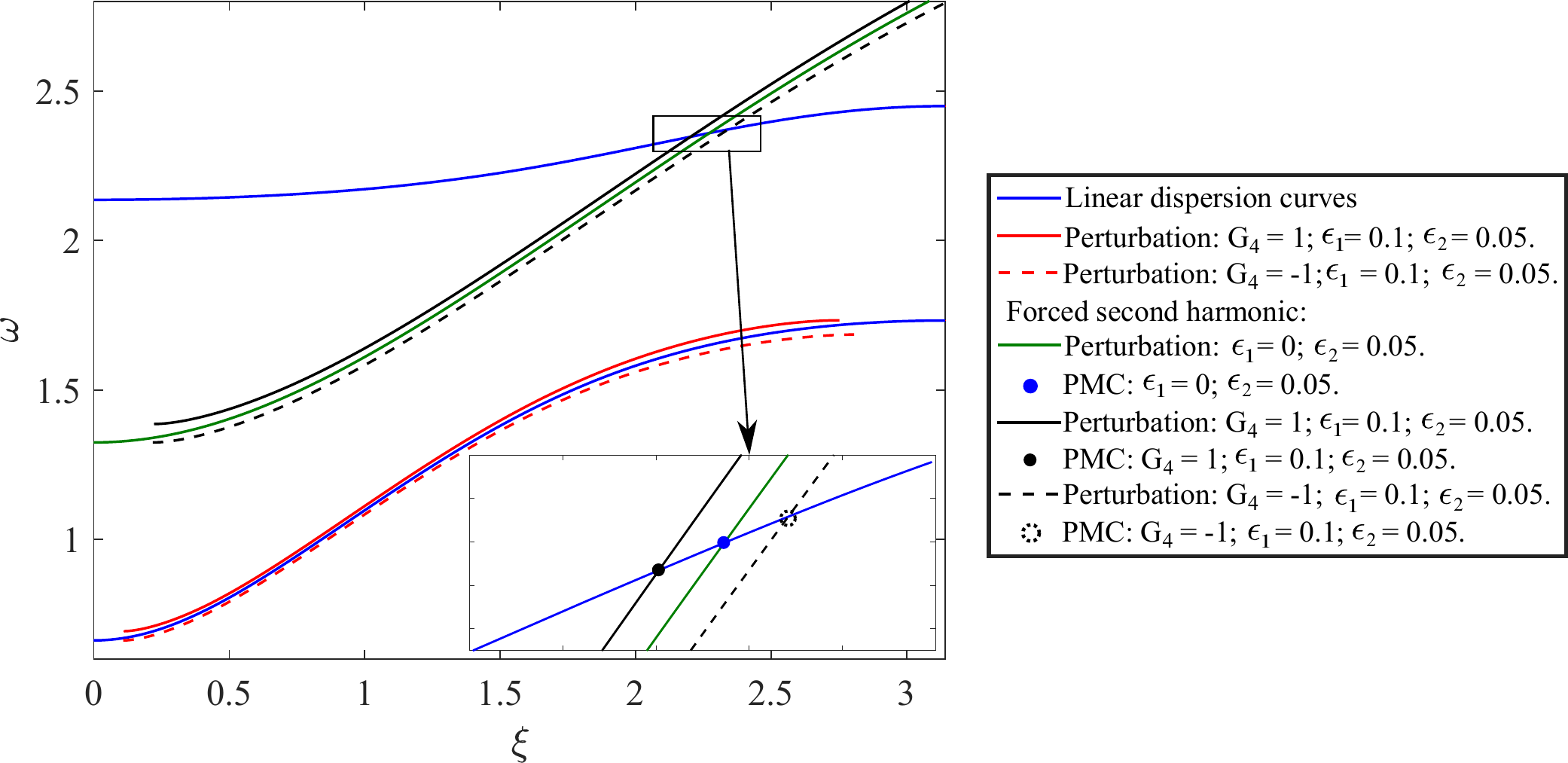}
	\caption{Schematic of tunable PMCs with hardening or softening cubic nonlinearity at amplitude $A_0=1$. }
	\label{Aplication2_illustration}
\end{figure*}

The second example exploits the possibility to interfere with the establishment of PMC in the second harmonic through the cubic correction. Since the strength of the harmonic generation under PMC is largely magnified, the entire modal manifestation of the second harmonic can be dramatically changed depending on whether PMC is enabled or not. 

\begin{figure*} [!htb]
	\centering
	\includegraphics[scale=0.6]{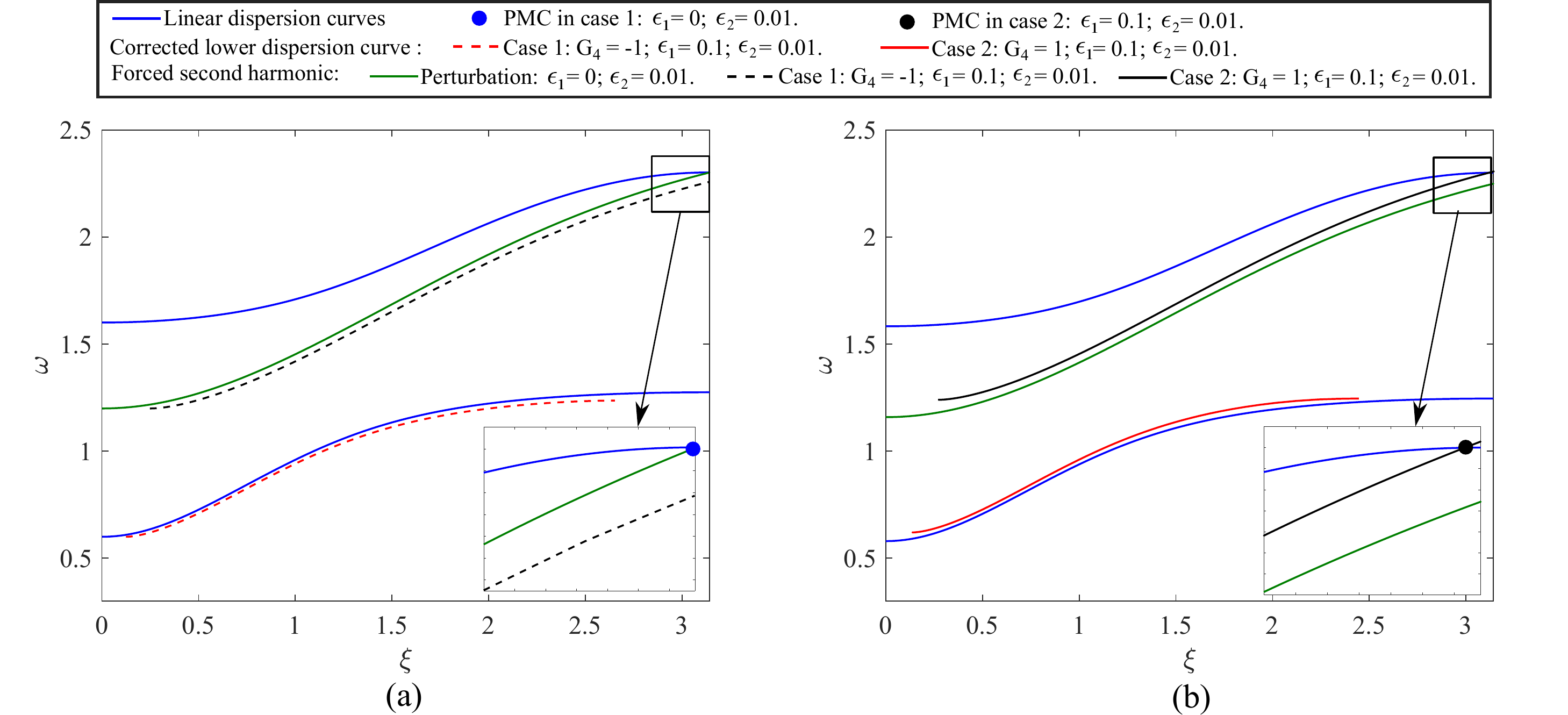}
	\caption{Schematic of self-adaptive PMC switches: (a) Case 1: switching \textit{off} PMC at high amplitude via hardening cubic nonlinearity; (a) Case 2: switching \textit{on} PMC at high amplitude via softening cubic nonlinearity.}
	\label{Aplication2_illustration_all}
\end{figure*}

As discussed in the theory section, if PMC is achieved, the amplitude of the solution at $O(\epsilon_2)$ is proportional to the propagation distance. This unique feature can be utilized to boost SHG, overall emphasizing the nonlinear response. However, for a given system with only quadratic nonlinearity, PMC is strictly linked to specific frequencies, or even not available at all, based on the fixed characteristics of the band diagram. In contrast, by incorporating cubic nonlinearity in the system and by exploiting its manifestation on the correction of the dispersion branches, we can instead switch on/off PMC in a certain frequency range, as shown in Fig.~\ref{Aplication2_illustration}. Based on its mathematical definition, the geometrical interpretation of PMC is the existence of an intersection between the linear dispersion curves and the curve encompassing all the forced second harmonic pair (the $2\tilde{\xi} - 2\omega$ curve). In light of this, in the system with $\epsilon_1=0$ blue{(i.e., no cubic nonlinearity)} and $\epsilon_2=0.05$ depicted in Fig.~\ref{Aplication2_illustration} PMC is located at the blue dot (zoomed details shown in the inset). If some hardening cubic nonlinearity ($\epsilon_1=0.1$, $G_4=1$) is added, PMC is shifted to the black dot. In contrast, if softening cubic nonlinearity is added ($G_4=-1$), PMC are shifted to the dashed black circle. In essence, by resorting to different types of cubic nonlinearity, PMC becomes highly tunable, providing another way to manipulate nonlinear wave propagation at the second harmonic. 

To illustrate this capability, we consider two chains. The parameters of chain 1 (with softening foundation) are $m_1=m_2=1$, $K_2=J_2=1=1$, $G_2=0.92$, $G_4=-1$, $\epsilon_1=0.1$, and $\epsilon_2=0.01$. Chain 2 (with hardening foundation) features $G_2=0.84$, $G_4=1$, while all the other coefficients are kept identical. For amplitude $A_0=1$, the corrected dispersion relations for both systems are plotted in Fig.~\ref{Aplication2_illustration_all} and superimposed to the linear ones. In the first case, the system features one PMC (blue dot in the inset of Fig.~\ref{Aplication2_illustration_all}(a)) close to the upper cut-off frequency, when excited with low amplitude or without cubic nonlinearity. By increasing the amplitude, the PMC can be switched off due to the activated dispersion shift (black dashed curve).  In the second case, there is no PMC available at low amplitude, but a new one (i.e., black dot in Fig.~\ref{Aplication2_illustration_all}(b)) can be created as we increase the amplitude to a certain degree and produce the correction marked by the black solid curve. 

\begin{figure*} [!htb]
	\centering
	\includegraphics[scale=0.4]{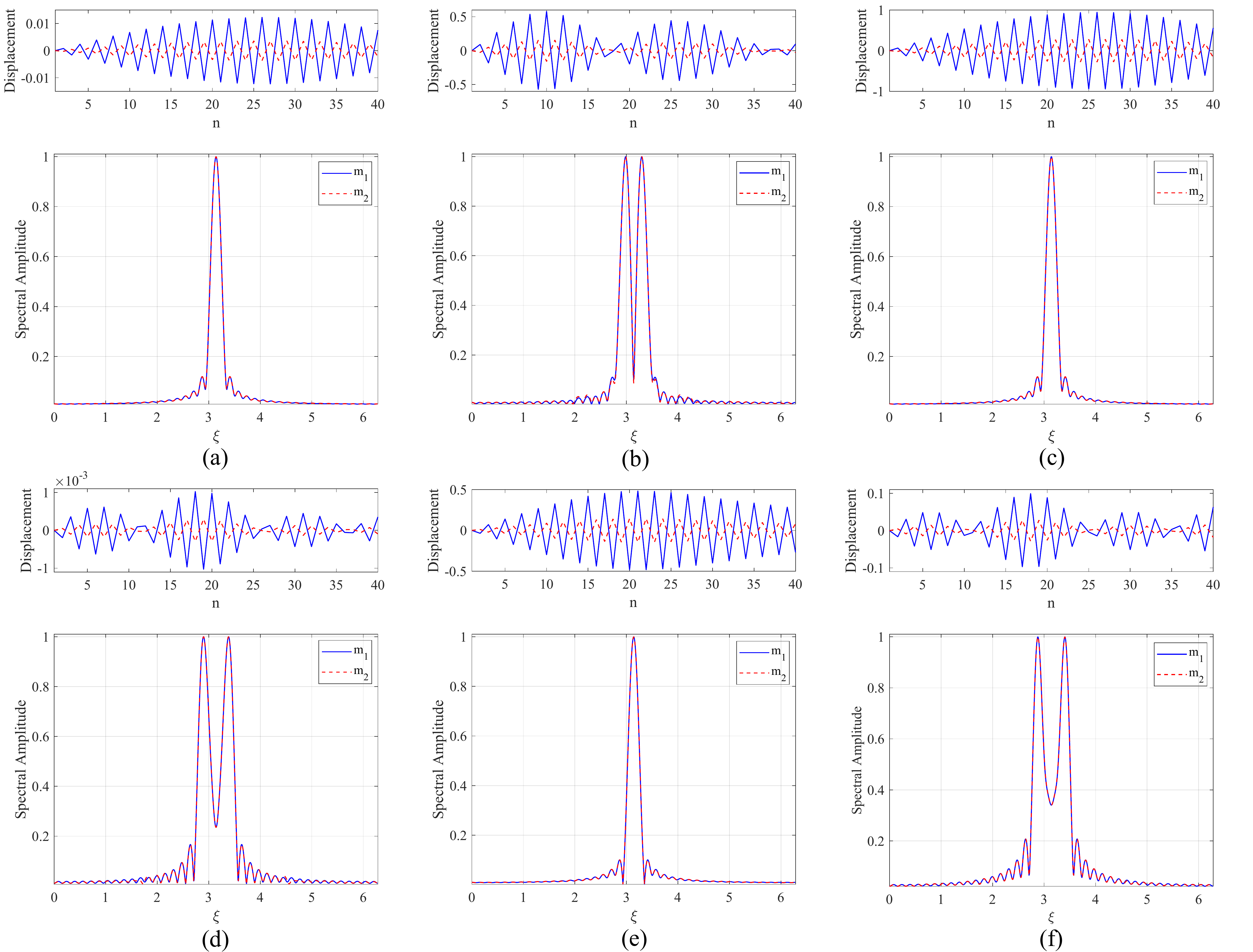}
	\caption{Spatial profile and DFT of the filtered second harmonics for the two cases of self-adaptive PMC switch. Case 1: (a) Low amplitude $A_0=0.1$ with $G_4=1$, $\epsilon_1=0.1$, $\epsilon_2=0.01$.  (b) High amplitude $A_0=1$ with $G_4=1$, $\epsilon_1=0.1$, $\epsilon_2=0.01$. (c) High amplitude $A_0=1$ with  $\epsilon_2=0.01$ (no cubic nonlinearity). Case 2: (d) Low amplitude $A_0=0.1$ with $G_4=-1$, $\epsilon_1=0.1$, $\epsilon_2=0.01$.  (e) High amplitude $A_0=1$ with $G_4=-1$, $\epsilon_1=0.1$, $\epsilon_2=0.01$. (f) High amplitude $A_0=1$ with $\epsilon_2=0.01$ (no cubic nonlinearity).}
	\label{Aplication2_simulation}
\end{figure*}

Next, we perform numerical simulations for both chains. The filtered second harmonics in the spatial and spectral domains are plotted in Fig.~\ref{Aplication2_simulation}. For chain 1, the excitation frequency is carefully chosen such that the second harmonic is close to the blue dot. At low excitation amplitude ($A_0=0.1$), PMC is achieved as designed, as confirmed by the fact that the spectrum only contains one dominant wavenumber component, as shown in Fig.~\ref{Aplication2_simulation}(a). As a result, the response at the second harmonic grows in space, as predicted from the analytical solution. Note however that, in practice, the harmonic starts losing intensity beyond a certain stage in the simulation. This reduction in amplitude is not captured by our perturbation analysis, because it occurs when the second harmonic, under PMC, is accumulated to a higher order than $O(\epsilon_2)$, which violates the order assumption in Eq.~\ref{soln_assumed}. At high amplitude ($A_0=1$), the PMC is switched off, and the second harmonic splits into two parts, one corresponding to the homogeneous response and the other denoting the forced response, which globally result a modulation in amplitude, as shown in Fig.~\ref{Aplication2_simulation}(b). In comparison, the result obtained with high amplitude of excitation but without additional cubic nonlinearity is given in Fig.~\ref{Aplication2_simulation}(c). We see that PMC is again enabled and the response features wave characteristics similar to those observed for the case of Fig.~\ref{Aplication2_simulation}(a). 

In the second case, the excitation frequency is chosen such that the second harmonic is close to the black dot, and therefore PMC is not available at low amplitude. As a result, the response at the second harmonic has two components as shown in the DFT plot of Fig.~\ref{Aplication2_simulation}(d). As we increase the amplitude to $1$, the dispersion shift leads to a nonlinearly-generated PMC, and the corresponding quasi-monochromatic numerical result is plotted in Fig.~\ref{Aplication2_simulation}(e). However, if we remove the cubic nonlinearity in the system, PMC is again unavailable, as confirmed by the resulting modulated filtered second harmonic shown in Fig.~\ref{Aplication2_simulation}(f). Similar to the previous example, the chain behaves like a self-adaptive switch, which can turn on and off selected modal characteristics typical of the optical mode via the interplay between cubic nonlinearity and SHG. Unlike the other case, however, here the mechanism is not the selective excitation and suppression of one component (homogeneous) of the second harmonic, but rather the selective activation of PMC and of its amplitude-reinforcing effects.

\section{Conclusion}

In this paper, we have studied wave propagation in a quadratic nonlinear spring-mass chain attached to a locally-resonant cubic nonlinear foundation, using a multiple scales analysis and full-scale simulations. We have been able to link the interplay between quadratic and cubic nonlinearity to a number of new nonlinear phenomena, as well as to revisit some more classical tuning and manipulation effects from a different perspective. 

First, we have revisited the amplitude-dependent dispersion shift of the fundamental harmonic, which has been known to be the most tangible effect of cubic nonlinearity. Here, we have shown that this effects has significantly different manifestations according to whether initial conditions or boundary conditions are prescribed. We have analyzed with special care the boundary conditions case, given its natural occurrence in practical problems of engineering interest. 

We have then considered the phenomenon of second harmonic generation, which is the prominent signature of quadratic nonlinearity. We have shown that, similarly to the fundamental harmonic, both homogeneous and forced second harmonics also feature amplitude-dependent dispersion shifts due to the Q-C nonlinearity interplay. 
We have presented two examples of tunable chains with self-adaptive switching capabilities. The first example involves activating or deactivating new modal characteristics in the nonlinear response through the dispersion shift induced in the homogeneous second harmonic. The second example shows the capability of switching on and off phase matching conditions to  activate and deactivate strong nonlinear signatures at high frequencies. In conclusion, the self-interacting effect of cubic and quadratic nonlinearities provides a new conceptual platform for the design of nonlinear phononic crystals and metamaterials with superior tunability characteristics.

\section*{Acknowledgements}
The authors acknowledge the support of the National Science Foundation (CAREER Award CMMI-$1452488$). 

\vspace{0.4cm}
\appendix* 
\section{Derivation of $f_u^2$}
The complete expression for the non-zero forcing term $f_u^2$ at $O(\epsilon_2)$, due to the quadratic nonlinearity, is
\begin{widetext}
	\begin{align}\label{A1}
	\begin{split}
	f_u^2&= -K_3\left[ \left(u^0_n-u^0_{n-1} \right) ^2 - \left(u^0_{n+1}-u^0_n \right) ^2\right] \\
	&= -K_3\left[A^2_0 \phi_u^2 \left(1-e^{-i\tilde{\xi}} \right)^2 e^{2i\tilde{\theta}_n}+c.c. + \underline{2A^2_0 \abs{\phi_u}^2 \left(1-e^{-i\tilde{\xi}} \right)\left(1-e^{i\tilde{\xi}} \right)}\right] \\ &+K_3\left[A^2_0 \phi_u \left(e^{i\tilde{\xi}}-1 \right)^2 e^{2i\tilde{\theta}_n} +c.c. +\underline{2A^2_0 \abs{\phi_u}^2 \left(e^{-i\tilde{\xi}}-1 \right)\left(e^{i\tilde{\xi}}-1 \right)}\right] \\
	&=-K_3A_0^2\phi_u^2\left[\left(1-e^{-i\tilde{\xi}} \right)^2 - \left(e^{i\tilde{\xi}}-1 \right)^2 \right] e^{2i\tilde{\theta}_n}+c.c. \\
	&=-8iK_3A^2_0 \phi_u^2\sin \tilde{\xi} \sin^2 \frac{\tilde{\xi}}{2} e^{2i\tilde{\theta}_n}+c.c.
	\end{split}
	\end{align}
\end{widetext}
\bibliography{myrefs} 

\end{document}